\documentclass[a4paper,amsmath,amssymb,nofootinbib,showpacs,notitlepage,byrevtex,showkeys,prd
]{revtex4-1}

\usepackage{graphicx}
\usepackage{subfigure}
\usepackage{bbm}

\newcommand{\vx}{\vec{x}}
\newcommand{\vz}{\vec{z}}

\newcommand{\vp}{\vec{p}}
\newcommand{\vk}{\vec{k}}
\newcommand{\vA}{\vec{A}}
\newcommand{\vB}{\vec{B}}
\newcommand{\vD}{\vec{D}}
\newcommand{\vy}{\vec{y}}
\newcommand{\vq}{\vec{q}}

\newcommand{\Id}{ \mathbbm{1}}
\renewcommand{\vec}[1]{\boldsymbol{#1}}

\newcommand{\beq}{\begin{equation}}
\newcommand{\eeq}{\end{equation}}
\newcommand*{\deltabar}{\delta\mkern-8mu\mathchar'26}
\newcommand{\ii}{\mathrm{i}}
\newcommand{\dd}{\mathrm{d}}
\newcommand*{\da}[1][]{\mathop{\mathrm{d}\mkern-7mu\mathchar'26\mkern-1mu^{#1}}\mkern-4mu}

\newcommand{\Dc}{\mathcal{D}}
\newcommand{\Nc}{\mathcal{N}}
\newcommand{\Oc}{\mathcal{O}}

\newcommand{\tr}{\mathrm{tr}}
\newcommand{\Ncc}{N_{\mathrm{C}}}

\newcommand{\vPi}{\mathbf{\Pi}}
\newcommand{\valpha}{\boldsymbol{\alpha}}

\newcommand{\hp}{\hat{\vp}}
\newcommand{\hq}{\hat{\vq}}
\newcommand{\hk}{\hat{\vk}}

\newcommand{\hpq}{\widehat{(\vp + \vq)}}

\usepackage[font=small]{caption}
\begin{document}

\title{Improved variational approach to QCD in Coulomb gauge}

\author{P.~Vastag, H.~Reinhardt and D.~Campagnari}
\affiliation{Institut f\"ur Theoretische Physik\\
Auf der Morgenstelle 14\\
D-72076 T\"ubingen\\
Germany}
\date{\today}

\begin{abstract}
The variational approach to QCD in Coulomb gauge developed previously by the T\"ubingen group is improved by enlarging the space of quark trial vacuum wave functionals through a new Dirac structure in the quark-gluon coupling. Our ansatz for the quark vacuum wave functional ensures that all linear divergences cancel in the quark gap equation resulting from the minimization of the energy calculated to two-loop order. The logarithmic divergences are absorbed in a renormalized coupling which is adjusted to reproduce the phenomenological value of the quark condensate. We also unquench the gluon propagator and show that the unquenching effects are generally small and amount to a small reduction in the mid-momentum regime.
\end{abstract}
\maketitle

\section{Introduction}

During the last two decades the infrared sector of QCD was intensively studied both on the lattice and in the continuum theory. Substantial insight into the two basic features of the QCD vacuum, confinement and spontaneous breaking of chiral symmetry (SBCS), has been gained although a rigorous understanding of these two phenomena is still lacking. Confinement is a property of the Yang--Mills sector. Indeed, the order parameter of confinement, the Wilson loop, is a purely gluonic observable, from which the static quark potential can be extracted. SBCS takes place in the quark sector: The quarks in the Dirac vacuum condense similar as the electrons in a superconductor, and the first microscopic explanations of the SBCS were based on superconductor types of effective models like the Nambu--Jona-Lasinio model \cite{NJL1961, NJL1961a, ER1986}. Indeed, the order parameter of SBCS is the quark condensate. However, by the Banks--Casher relation \cite{BC1980} this order parameter is related to the level density near zero virtuality of the quarks in the fluctuating gluonic background. Consequently, SBCS is also caused indirectly by the gluons.

Extensive lattice studies have shown that confinement and SBCS are both caused by topologically non-trivial field configurations like center vortices and magnetic monopoles, see ref.~\cite{Greensite2011} for a review. Indeed, when center vortices or magnetic monopoles are removed by hand from the ensemble of gauge field configurations the area law of the Wilson loop and thus the confining part of the extracted quark potential is lost \cite{DelDebbio1998, ELRT2000}. At the same time the level density of the quarks develops a gap near zero virtuality \cite{Gattnar2005} and, by the Banks--Casher relation, SBCS disappears. On the other hand, when one projects the ensemble of gauge field configurations on those containing center vortices, the short distance Coulomb part of the static quark potential disappears while the linearly rising confining part is preserved at all distances. At the same time the quark levels are squeezed in the region around zero virtuality \cite{Hoellwieser2008}.

A seemingly different confinement scenario was proposed by Gribov \cite{Gribov1978} and further developed by Zwanziger \cite{Zwanziger1998}, in which confinement manifests itself in an infrared diverging ghost form factor. Extensive studies, both on the lattice and in the continuum, have shown that this picture is not realized in Landau gauge \cite{CM2007} (as originally assumed) but only in Coulomb gauge \cite{Feuchter2004, Feuchter2004a, ERS2007, BQR2009}. Though confinement is a gauge invariant phenomenon, it may manifest itself differently in different gauges. In Coulomb gauge not only an infrared diverging ghost form factor but also a linearly rising static quark potential (the so-called non-Abelian Coulomb potential) is obtained. The infrared slope of this potential is given by the so-called Coulomb string tension, which is an upper bound to the Wilson string tension \cite{Zwanziger2003}. The Gribov--Zwanziger picture was mainly established within the Hamiltonian approach to QCD in Coulomb gauge by means of variational calculations \cite{Feuchter2004, Feuchter2004a, ERS2007, SS2001} and is also supported by lattice calculations \cite{BQR2009, GOZ2004}. Lattice and continuum studies have shown that the Gribov--Zwanziger confinement scenario is tightly related to the center vortex and magnetic monopole pictures of confinement: Center vortices and magnetic monopoles live on the so-called Gribov horizon \cite{GOZ2004}, i.e.~are field configurations for which the Faddeev--Popov determinant vanishes. When these center vortices are removed from the Yang--Mills ensemble, the ghost form factor becomes infrared finite and the non-Abelian Coulomb potential is no longer linearly rising but becomes infrared flat \cite{GOZ2004}, i.e.~the Coulomb string tension disappears. Recently, it was also shown that the Coulomb string tension is not related to the temporal Wilson string tension but to the spatial string tension \cite{BQRV2015}. This also explains why the Coulomb string tension does not disappear above the deconfinement phase transition \cite{VBQR2014}. Furthermore, within the Hamiltonian approach in Coulomb gauge it was shown that the inverse of the ghost form factor can be interpreted as the dielectric function of the Yang--Mills vacuum \cite{Reinhardt2008}. An infrared diverging ghost form factor then implies that the dielectric function vanishes in the infrared, which makes the Yang--Mills vacuum a perfect color dielectricum, i.e.~a dual superconductor, which arises from the condensation of magnetic monopoles. In this sense the Gribov--Zwanziger picture is also related to the magnetic monopole picture of confinement.

Variational calculations within the Hamiltonian approach in Coulomb gauge were initiated in ref.~\cite{Schuette1984}, where a Gaussian trial ansatz was used for the Yang--Mills vacuum wave functional. The same ansatz was used in ref.~\cite{SS2001} where also the first numerical calculations were carried out. The approach developed by the T\"ubingen group \cite{Feuchter2004, Feuchter2004a, ERS2007} differs from previous work in the choice of the trial wave functional and more importantly in the treatment of the Faddeev--Popov determinant as well as in the renormalization, see ref.~\cite{Greensite2011a} for more details. In fact, in previous work the Faddeev--Popov determinant was not (properly) included. However, it turns out that the Faddeev--Popov determinant is crucial for the infrared properties of the theory and for the Gribov--Zwanziger picture to be realized \cite{Feuchter2004, Feuchter2004a, ERS2007}. Within our approach we have obtained a decent description of the infrared properties of QCD as, for example, an infrared diverging gluon energy \cite{Feuchter2004, Feuchter2004a, ERS2007} (which is a signal of gluon confinement and also supported by the lattice calculation \cite{BQR2009}), a linearly rising non-Abelian Coulomb potential, an infrared finite running coupling constant \cite{ERS2007}, a perimeter law for the 't Hooft loop \cite{RE2007} and an area law for the Wilson loop \cite{PR2009}. For a recent review, see ref.~\cite{Reinhardt2015}.

The advantage  of the Hamiltonian approach in Coulomb gauge is that the gauge fixed Hamiltonian contains  already a confining four-quark interaction, which depends on the fluctuating transversal gauge field by the so-called Coulomb kernel, see eqs.~(\ref{Gl: Coulombterm}) and (\ref{Gl: Coulombkern}). If this kernel is replaced by its Yang--Mills vacuum expectation value a confining quark potential is obtained. Keeping from the gauge fixed QCD Hamiltonian only this confining two-body interaction and the free Dirac Hamiltonian of the quarks one obtains a confining quark model, which has been treated originally by Finger and Mandula within a variational approach using for the quark vacuum wave functional a BCS-type of ansatz \cite{FM1982}. From this model one finds indeed SBCS but the quark condensate turns out to be much too small compared to the phenomenological values when realistic values for the Coulomb string tension are used. The model was reconsidered in \cite{Adler1984} where the renormalization was improved and also in ref.~\cite{LeYaouanc1984} and was extended to non-zero current quark masses in ref.~\cite{AA1988}. Similar studies of chiral symmetry breaking within quark models with confining two-body interactions were carried out in \cite{BR1990} and \cite{SK2002}.

The variational approach to Yang--Mills theory developed in ref.~\cite{Feuchter2004, Feuchter2004a} was extended in refs.~\cite{Pak2012a, Pak2013} to full QCD. For the quark vacuum wave functional a trial ansatz was used, which goes beyond the BCS-type of state considered previously in refs.~\cite{FM1982, Adler1984, AA1988} by explicitly including the coupling of the quarks to the spatial gluons. It was shown that the inclusion of the quark-gluon coupling substantially increases the amount of chiral symmetry breaking. Unfortunately, in ref.~\cite{Pak2012a} a simplifying approximation was used in the evaluation of the gluonic expectation value of quark observables, which leads to the unrealistic property that the form factor of the quark-gluon coupling term in the wave functional depends only on one independent momentum, while on general grounds, with the overall momentum conservation taken into account, it should depend on two independent momenta. In the present paper we will go beyond ref.~\cite{Pak2013} and develop an improved variational approach to QCD in Coulomb gauge. The improvement is twofold: First, we will use a generalized ansatz for the quark vacuum wave functional, which (compared to ref.~\cite{Pak2013}) includes an additional quark-gluon coupling term with a new Dirac structure. This term can be motivated by perturbation theory and has the advantage that it removes all linear UV divergences from the quark gap equation. Second, we will abandon the approximation used in ref.~\cite{Pak2013} and calculate the expectation value of the QCD Hamiltonian consistently to two-loop order.

The organization of this paper is as follows: In the next section we present the main features of the Hamilton formulation of QCD in Coulomb gauge and fix our notation. In section \ref{Abschn: YM} we summarize the essential results obtaind within the variational approach in Coulomb gauge for the Yang--Mills sector, which will serve as input for the quark sector. Our trial ansatz for the quark vacuum wave functional is presented in section \ref{Abschn: Quark}. The quark propagator is calculated in section \ref{Abschn: Propagator} for our trial wave functional and in section \ref{Abschn: EW} the vacuum energy is determined to two-loop order. The equations of motion for the variational kernels of our vacuum wave functional are derived in section \ref{Abschn: Variation} by minimizing the energy. The UV analysis of these equations and their renormalization are carried out in section \ref{Abschn: UV}. In section \ref{Abschn: Skalierung} we study the physical implications of the coupling of the quarks to the spatial gluons. Our numerical results are presented in section \ref{Abschn: Numerik}. A short summary and our conclusions are given in section \ref{Abschn: Summary}. Some mathematical details are presented in appendices.

\section{Hamiltonian formulation of QCD in Coulomb gauge} \label{Abschn: Hamiltonzugang}

Canonical quantization of QCD in Weyl, $A_0 = 0$, and Coulomb gauge, $\nabla \cdot \vA = 0$, results in the following Hamiltonian \cite{Christ1980, Pak2012a, Pak2013}
\beq
H_{\mathrm{QCD}} = H_{\mathrm{YM}} + H_{\mathrm{Q}} + H_{\mathrm{C}} \, . \label{Gl: QCDHamiltonian}
\eeq
Here
\beq
H_{\mathrm{YM}} = \frac{1}{2} \int \dd^3 x \left(J^{-1}[A] \, \vPi(\vx) J[A] \, \vPi(\vx) + \vB^2(\vx) \right) \equiv H^E_{\mathrm{YM}} + H^B_{\mathrm{YM}} \label{Gl: YMHamiltonian}
\eeq
is the gauge fixed Hamiltonian of the transversal components of the gauge field $A$ which satisfy $A_k = t_{kl} A_l$ with the transversal projector \mbox{$t_{kl}(\vx) = \delta_{kl} + \partial_k (-\Delta)^{- 1} \partial_l$}. In eq.~(\ref{Gl: YMHamiltonian})
\beq
B^a_k(\vx) = \varepsilon_{klm} \left(\partial_l A^a_m(\vx) - \frac{g}{2} f^{abc} A^b_l(\vx) A^c_m(\vx)\right) \label{Gl: Farbmagnetfeld}
\eeq
is the non-Abelian color magnetic field with bare coupling constant $g$ and structure constants $f^{a b c}$ of the color group. Furthermore, 
\beq
\Pi^a_k(\vx) = \frac{\delta}{\ii \delta A^a_k(\vx)} \label{Gl: KanImpuls}
\eeq
is the canonical momentum operator conjugate to the transversal gauge field in coordinate representation. It represents the operator of the color electric field and fulfills the canonical commutator relations
\begin{subequations}
\begin{align}
\left[A_k^a(\vx), \Pi_l^b(\vy)\right] &= \ii \delta^{a b} t_{k l}(\vx - \vy) \\
\left[A_k^a(\vx), A_l^b(\vy)\right] &= \left[\Pi_k^a(\vx), \Pi_l^b(\vy)\right] = 0 \,.
\end{align}
\end{subequations}
Finally,
\beq
J[A] = \det\bigl({\hat{G}}^{-1}\bigr) \label{Gl: FaddeevPopov}
\eeq
is the Faddeev--Popov determinant in Coulomb gauge, where
\beq
\bigl(\hat{G}^{-1}\bigr)^{a b}(\vx, \vy) \equiv \bigl(-\nabla \cdot \hat{\vD}\bigr)^{a b}(\vx, \vy)
\eeq
is the Faddeev--Popov operator with
\beq
\hat{D}^{a b}_k(\vx) = \delta^{ab} \partial_k - g f^{acb} A^c_k(\vx) \label{Gl: KovAbleitung}
\eeq
being the covariant derivative in the adjoint representation.

The second term in eq.~(\ref{Gl: QCDHamiltonian}),
\beq
H_{\mathrm{Q}} = \int \dd^3 x \, \psi^\dagger(\vx) \Bigl[\vec{\alpha} \cdot \bigl(-\ii \nabla + g t^a \vA^a(\vx)\bigr) + \beta m_{\mathrm{Q}} \Bigr] \psi(\vx) \equiv H_{\mathrm{Q}}^0 + H_{\mathrm{Q}}^A \, , \label{Gl: DiracHamiltonian}
\eeq
is the Hamiltonian of the quarks in the background of the fluctuating gauge field $A$. The quark field $\psi$ satisfies the usual equal time anti-commutation relation
\begin{subequations}
\begin{align}
\Bigl\{\psi_i^m(\vx), {\psi_j^n}^\dagger(\vy)\Bigr\} &= \delta^{m n} \delta_{i j} \delta (\vx - \vy) \\
\Bigl\{\psi_i^m(\vx), \psi_j^n(\vy)\Bigr\} &= 0 \, .
\end{align}
\label{Gl: Antikommutator}%
\end{subequations}
Furthermore, $\alpha$, $\beta$ in eq.~(\ref{Gl: DiracHamiltonian}) are the usual Dirac matrices and $t^a$ denotes the generator of the color group in the fundamental representation. Finally, $m_{\mathrm{Q}}$ is the bare quark mass. In this paper we will consider only one quark flavor but the extension to several flavors with different quark masses is straightforward.

Finally, the last term in eq.~(\ref{Gl: QCDHamiltonian}) is the so-called Coulomb term, which arises from the longitudinal part of the gluonic kinetic energy after resolution of Gau{\ss}'s law in Coulomb gauge and is given by 
\beq
H_{\mathrm{C}} = \frac{g^2}{2} \int \dd^3 x \int \dd^3 y \, J^{- 1}[A] \rho^a(\vx) J[A] \hat{F}^{ab}(\vx, \vy) \rho^b(\vy) \, , \label{Gl: Coulombterm}
\eeq
where the Coulomb kernel
\beq
\hat{F}^{a b}(\vx, \vy) = \int \dd^3 z \, \hat{G}^{a c}(\vx, \vz) (-\Delta_{\vz}) \hat{G}^{c b}(\vz, \vy) \label{Gl: Coulombkern}
\eeq
is a highly non-local functional of the gauge field. Furthermore,
\beq
\rho(\vx) = \rho_{\mathrm{YM}}(\vx) + \rho_{\mathrm{Q}}(\vx) \label{Gl: Farbdichte}
\eeq
is the total color charge density, which besides the quark part
\beq
\rho^a_{\mathrm{Q}}(\vx) = \psi^\dagger(\vx) t^a \psi(\vx) \label{Gl: FarbdichteQ}
\eeq
receives also a gluonic contribution given by
\beq
\rho^a_{\mathrm{YM}}(\vx) = f^{abc} \vA^b(\vx) \cdot \vPi^c(\vx) \, . \label{Gl: FarbdichteYM}
\eeq
Note that the gluonic charge density (\ref{Gl: FarbdichteYM}) does not commute with the Faddeev--Popov determinant.

Since the total color charge density $\rho$ (\ref{Gl: Farbdichte}) is the sum of a quark and a gluon part, the Coulomb Hamiltonian (\ref{Gl: Coulombterm}) can be expressed as
\beq
H_{\mathrm{C}} = H_{\mathrm{C}}^{\mathrm{YM}} + H_{\mathrm{C}}^{\mathrm{INT}} + H_{\mathrm{C}}^{\mathrm{Q}} \, , \label{Gl: Coulombterm1}
\eeq
where $H_{\mathrm{C}}^{\mathrm{YM}}$ and $H_{\mathrm{C}}^{\mathrm{Q}}$ depend exclusively on the charges of the gauge field, $\rho_{\mathrm{YM}}$, and the quark field, $\rho_{\mathrm{Q}}$, respectively, while $H_{\mathrm{C}}^{\mathrm{INT}}$ contains the coupling between both. Note that the Faddeev--Popov determinant drops out from the quark part
\beq
H_{\mathrm{C}}^{\mathrm{Q}} = \frac{g^2}{2} \int \dd^3 x \int \dd^3 y \, \rho_{\mathrm{Q}}^a(\vx) \hat{F}^{a b}(\vx, \vy) \rho_{\mathrm{Q}}^b(\vy) \, . \label{Gl: CoulombtermQ}
\eeq
For subsequent considerations it will be convenient to reshuffle the QCD Hamiltonian as
\beq
H_{\mathrm{QCD}} = \bar{H}_{\mathrm{YM}} + \bar{H}_{\mathrm{Q}} \, , \label{Gl: QCDHamiltonian1}
\eeq
where
\beq
\bar{H}_{\mathrm{YM}} = H_{\mathrm{YM}} + H_{\mathrm{C}}^{\mathrm{YM}} \label{Gl: YMHamiltonian1}
\eeq
is the Hamiltonian of the pure Yang--Mills sector (i.e. $\bar{H}_{\mathrm{YM}}$ does not contain the quark field) and
\beq
\bar{H}_{\mathrm{Q}} = H_{\mathrm{Q}} + H_{\mathrm{C}}^{\mathrm{Q}} + H_{\mathrm{C}}^{\mathrm{INT}} \label{Gl: QuarkHamiltonian}
\eeq
decribes the quark sector coupled to the gluons.

The aim of the Hamiltonian approach is to solve the functional Schr\"odinger equation
\beq
H_{\mathrm{QCD}} \vert \phi \rangle = E \vert \phi \rangle \label{Gl: Schroedinger}
\eeq
for the vacuum state of QCD. In the present paper we attempt this in an approximate fashion by exploiting the variational principle. In analogy to the splitting (\ref{Gl: QCDHamiltonian1}) of the QCD Hamiltonian we write the QCD vacuum wave functional in the factorized form
\beq
\vert \phi[A] \rangle = \phi_{\mathrm{YM}}[A] \, \vert \phi_{\mathrm{Q}}[A] \rangle \, , \label{Gl: Vakuumfunktional}
\eeq 
where $\vert \phi_{\mathrm{Q}}[A] \rangle$ is the wave functional of the Dirac sea of the quarks in the background of the fluctuating gauge field $A$ and $\phi_{\mathrm{YM}}[A] = \langle A \vert \phi_{\mathrm{YM}} \rangle$ is the vacuum wave functional of the Yang--Mills sector. The ansatz (\ref{Gl: Vakuumfunktional}) is, in principle, exact, since the quark wave functional $\vert \phi_{\mathrm{Q}}[A] \rangle$ depends explicitly on the gluon field, and thus could capture the entire quark-gluon interaction. Of course, in the actual calculation we have to restrict the trial wave functionals to a subspace of the whole Hilbert space. Note also that we have chosen the coordinate representation for the Yang--Mills part of the vacuum wave functional while for the fermionic part we prefer to use the second quantized form of the Fock space representation. In principle, we could also use a coordinate representation for the quark wave functional, which is defined in terms of Gra{\ss}mann variables \cite{CR2015}. However, the mixed representation (\ref{Gl: Vakuumfunktional}) turns out to be not only sufficient but also quite convenient.\footnote{However, when the variational approach is formulated for non-Gaussian trial states by means of the generalized Dyson--Schwinger equations, the use of the coherent fermion state basis of the Fock space in terms of  Gra{\ss}mann variables is essential.}

The expectation value of an observable $O[A, \Pi, \psi]$ in the state (\ref{Gl: Vakuumfunktional}) is given by 
\beq
\langle O[A, \Pi, \psi] \rangle = \int \Dc A \, J[A] \phi_{\mathrm{YM}}^*[A] \, \langle \phi_{\mathrm{Q}}[A] \vert O[A, \Pi, \psi] \vert \phi_{\mathrm{Q}}[A] \rangle \, \phi_{\mathrm{YM}}[A] \, , \label{Gl: EW}
\eeq
where the Faddeev--Popov determinant arises in the integration measure of the gauge field from the fixing to Coulomb gauge. In the present paper we will concentrate on the determination of the quark vacuum wave functional $\vert \phi_{\mathrm{Q}}[A] \rangle$ using the results obtained previously in the Yang--Mills sector as input. For this purpose, we will briefly summarize these results in the next section.

\section{Variational results for the gluon sector of QCD} \label{Abschn: YM}

\begin{figure}
\centering
\includegraphics[width=0.4\linewidth]{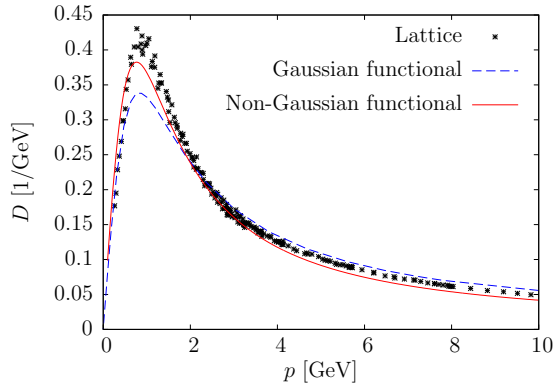}
\caption{The static gluon propagator $D(p) = 1 / 2 \omega(p)$ in momentum space. Crosses are lattice data, while the dashed and full lines, respectively, refer to the results of the variational approach with a Gaussian \cite{Feuchter2004, Feuchter2004a} and a non-Gaussian \cite{CR2010} vacuum wave functional.}
\label{Abb: Gluonpropagator}
\end{figure}

In refs.~\cite{Feuchter2004, Feuchter2004a, ERS2007} the gluon sector of QCD in Coulomb gauge was treated in a variational approach using the following ansatz for the vacuum state
\beq
\phi_{\mathrm{YM}}[A] = \Nc_{\mathrm{YM}} J^{-\frac{1}{2}}[A] \exp\left(-\frac{1}{2} \int \dd^3 x \int \dd^3 y \, A_k^a(\vx) \omega(\vx, \vy) A_k^a(\vy)\right) , \label{Gl: VakuumfunktionalYM}
\eeq
where $\Nc_{\mathrm{YM}}$ is a normalization factor fixed by the condition $\langle \phi_{\mathrm{YM}} \vert \phi_{\mathrm{YM}} \rangle = 1$, $J$ is the Faddeev--Popov determinant (\ref{Gl: FaddeevPopov}) and $\omega$ is a variational kernel. Compared to  a pure Gaussian, this ansatz with the preexponential factor included has the advantage that the Faddeev--Popov determinant drops out from the integration measure of the scalar product (\ref{Gl: EW}), and as a consequence the static gluon propagator is given by
\beq
D^{a b}_{k l} (\vx, \vy) = \langle A_k^a(\vx) A_l^b(\vy) \rangle_{\mathrm{YM}} = \frac{1}{2} \delta^{a b} t_{k l}(\vx) \omega^{-1}(\vx, \vy) \, , \label{Gl: Gluonpropagator}
\eeq
where $\langle \ldots \rangle_{\mathrm{YM}}$ denotes the expectation value in the Yang--Mills vacuum state $\phi_{\mathrm{YM}}$ [eq.~(\ref{Gl: VakuumfunktionalYM})]. Furthermore, the ansatz (\ref{Gl: VakuumfunktionalYM}) guarantees that Wick's theorem holds so that all gluonic expectation values $\langle \ldots \rangle_{\mathrm{YM}}$ can, in principle, be entirely expressed in terms of the gluon propagator (\ref{Gl: Gluonpropagator}). In ref.~\cite{Feuchter2004, Feuchter2004a} the energy of the gluon sector $\langle \bar{H}_{\mathrm{YM}} \rangle_{\mathrm{YM}}$ was calculated up to two loops. This implies that for the expectation value of the Coulomb kernel $\hat{F}$ [eq.~(\ref{Gl: Coulombkern})] the factorization
\beq
\langle \hat{G} (-\Delta) \hat{G} \rangle_{\mathrm{YM}} \approx \langle \hat{G} \rangle_{\mathrm{YM}} (-\Delta) \langle \hat{G} \rangle_{\mathrm{YM}} \label{Gl: Coulombkern1}
\eeq
was used, where $\langle \hat{G} \rangle_{\mathrm{YM}}$ is the ghost propagator. Furthermore, up to two-loop order in the energy it is sufficient to replace the Faddeev--Popov determinant by the Gaussian functional \cite{Feuchter2005}
\beq
J[A] = \exp\left(-\int \dd^3 x \int \dd^3 y \, A_k^a(\vx) \chi_{k l}^{a b}(\vx, \vy) A_l^b(\vy)\right) \label{Gl: FaddeevPopov1}
\eeq
where
\beq
\chi_{k l}^{a b}(\vx, \vy) = -\frac{1}{2} \Bigl\langle \frac{\delta}{\delta A_k^a(\vx)} \frac{\delta}{\delta A_l^b(\vy)} \ln J[A] \Bigr\rangle_{\mathrm{YM}} \label{Gl: Kruemmung}
\eeq
denotes the ghost loop, which in this context is referred to as curvature. Minimizing the energy with respect to $\omega$ results in a gluonic gap equation which contains only up to one-loop terms. This equation together with the Dyson--Schwinger equation for the ghost propagator can be solved analytically both in the infrared and the ultraviolet by power law ans\"atze. One finds that the gluon energy behaves like the photon energy, $\omega \sim p$, for large momenta $p \to \infty$, while it is infrared diverging, $\omega \sim 1/p$ for $p \to 0$. These analytic results are confirmed by the full numerical calculations which are compared in fig.~\ref{Abb: Gluonpropagator} with lattice data. The gluon propagator obtained in the variational approach \cite{Feuchter2004, Feuchter2004a} agrees nicely with the lattice data in the infrared and in the ultraviolet but misses some strength in the mid-momentum regime. In this regime the variational results can be considerable improved by using a non-Gaussian trial state which in the exponent includes also terms cubic and quartic in the gauge field \cite{CR2010}, see fig.~\ref{Abb: Gluonpropagator}. The most remarkable feature of the lattice data for the gluon propagator is that its representation eq.~(\ref{Gl: Gluonpropagator}) can be nicely fitted by the Gribov formula \cite{Gribov1978}
\beq
\omega(p) = \sqrt{p^2 + \frac{M_{\mathrm{G}}^4}{p^2}} \, \label{Gl: Gribov}
\eeq
with a Gribov mass of $M_{\mathrm{G}} \approx 880 \, \mathrm{MeV} \approx 2 \sqrt{\sigma}$ \cite{BQR2009}, where $\sigma$ is the Wilsonian string tension. From eq.~(\ref{Gl: CoulombtermQ}) it is seen that the gluonic vacuum expectation value of the Coulomb kernel $\hat{F}$ [eq.~(\ref{Gl: Coulombkern})]
\beq
g^2 \langle F^{a b}(\vx, \vy) \rangle_{\mathrm{YM}} = \delta^{a b} V_{\mathrm{C}}(|\vx - \vy|) \, \label{Gl: Coulombkern3}
\eeq
represents a static color charge potential. At small distances this potential can be calculated in perturbation theory and one finds
\beq
V_\mathrm{C}(r) \xrightarrow[r \to 0]{}  \frac{\alpha_{\mathrm{S}}}{r} \, , \quad \quad \alpha_{\mathrm{S}} = \frac{g^2}{4 \pi} \label{Gl: Coulombkern4}
\eeq
in agreement with asymptotic freedom. With the approximation (\ref{Gl: Coulombkern1}) it was found \cite{ERS2007} that at large distances this potential rises linearly,
\beq
V_{\mathrm{C}}(r) \xrightarrow[r \to \infty]{} \, \sigma_{\mathrm{C}} r \, , \label{Gl: Coulombkern5}
\eeq
with a coefficient $\sigma_{\mathrm{C}}$ referred to as Coulomb string tension. This quantity can be shown \cite{Zwanziger2003} to be an upper bound to the Wilson string tension $\sigma$ extracted from the Wilson loop. On the lattice one finds also a static non-Abelian Coulomb potential growing linearly in the infrared with $\sigma_{\mathrm{C}} = 2 \ldots 4 \, \sigma$ \cite{BQRV2015, GOZ2004, Voigt2008}. The infrared analysis of the variational equations of motion (gluon gap equation and ghost Dyson--Schwinger equation) \cite{ERS2007} shows that within the approximation (\ref{Gl: Coulombkern1}) the Coulomb string tension $\sigma_{\mathrm{C}}$ is related to the Gribov mass $M_{\mathrm{G}}$ by
\beq
\sigma_{\mathrm{C}} = \frac{\pi}{N_{\mathrm{C}}} M_{\mathrm{G}}^2 \, . \label{Gl: Coulombstringtension}
\eeq
For $N_{\mathrm{C}} = 3$ and $\pi / N_{\mathrm{C}} \approx 1$ we obtain $\sigma_{\mathrm{C}} \approx M_{\mathrm{G}}^2$, which shows that there is a single mass scale in the gluon sector. Furthermore, with the lattice result $M_{\mathrm{G}} \approx 2 \sqrt{\sigma}$ \cite{BQR2009} we obtain for the Coulomb string tension $\sigma_{\mathrm{C}} \approx 4 \sigma$, which is at the upper border of the range found for $\sigma_{\mathrm{C}}$ on the lattice. It is also worth mentioning that to the order of approximation considered the bare coupling constant $g$ drops out from the variational equations of motion of the gluon sector so that these equations are scale invariant and the physical scale has to be determined by calculating some physical observable. Following ref.~\cite{Heffner2012} we will use the Coulomb string tension $\sigma_{\mathrm{C}}$ measured on the lattice to fix the scale.

Finally, as shown in ref.~\cite{Heffner2012} the Coulomb term is completely irrelevant in the gluon sector, i.e. $H_{\mathrm{C}}^{\mathrm{YM}}$ can be 
safely neglected. Furthermore, the interaction term $H_{\mathrm{C}}^{\mathrm{INT}}$ contributes to $\langle H_{\mathrm{QCD}} \rangle$ only in higher than two-loop order and will hence be neglected. On the other hand, the quark part $H_{\mathrm{C}}^{\mathrm{Q}}$ (with the Coulomb kernel $\hat{F}$ replaced by its expectation value (\ref{Gl: Coulombkern3})) contributes already at the two-loop level and has to be kept. We will later find that this term is in fact quite important for the spontaneous breaking of chiral symmetry. The Hamiltonian of the quark sector then reads
\beq
\bar{H}_{\mathrm{Q}} = H_{\mathrm{Q}} + H_{\mathrm{C}}^{\mathrm{Q}} \label{Gl: QuarkHamiltonian1}
\eeq
with the Coulomb kernel $\hat{F}$ in $H_{\mathrm{C}}^{\mathrm{Q}}$ replaced by its gluonic vacuum expectation value $V_{\mathrm{C}}$ [eq.~(\ref{Gl: Coulombkern3})].

\section{The quark wave functional} \label{Abschn: Quark}

Our trial ansatz for the quark wave functional is an extension of that used in refs.~\cite{Pak2012a, Pak2013}. Following these references, we split the quark field operator $\psi$ into its positive and negative energy components
\beq
\psi(\vx) =  \psi_+(\vx) + \psi_-(\vx) \label{Gl: Feldoperatorzerlegung}
\eeq
defined with respect to the bare Dirac vacuum $\vert 0 \rangle$
\beq
\psi_+(\vx) \vert 0 \rangle = 0 = \psi_-^{\dagger}(\vx) \vert 0 \rangle \, . \label{Gl: DiracVakuum}
\eeq
The bare Dirac vacuum consists of the filled negative energy eigenstates of the bare Dirac Hamiltonian, which in momentum space reads
\beq
h(\vp) = \valpha \cdot \vp + \beta m_{\mathrm{Q}} \, \label{Gl: DiracOperator}
\eeq
and whose  eigenvalues are given by $\pm e(p) = \pm \sqrt{p^2 + m_{\mathrm{Q}}^2}$. The positive and negative energy components of the quark field can then be expressed as
\beq
\psi_{\pm}(\vx) = \int \dd^3 y \, \Lambda_{\pm}(\vx, \vy) \psi(\vy) \label{Gl: Feldoperatorzerlegung1}
\eeq
with the orthogonal projectors ($\da = \dd / (2 \pi)$)
\beq
\Lambda_{\pm}(\vx, \vy) = \int \da^3 p \, \exp\bigl(\ii \vp \cdot(\vx - \vy)\bigr) \Lambda_{\pm}(\vp) \, , \quad \quad \Lambda_{\pm} (\vp)
= \frac{1}{2} \left(\Id \pm \frac{h(\vp)}{e(p)}\right) \label{Gl: Projektor}
\eeq
satisfying
\beq
\Lambda_+ + \Lambda_- = \Id \, , \quad \quad \Lambda_{\pm}^2 = \Lambda_{\pm} \, , \quad \quad \Lambda_{\pm} \Lambda_{\mp} = 0 \, . \label{Gl: Projektor1}
\eeq
From (\ref{Gl: Antikommutator}) it follows that the projected quark fields $\psi_{\pm}$ obey the anti-commutation relations
\begin{subequations}
\begin{alignat}{2}
\left\{\psi_{\pm}(\vx), \psi_{\pm}^{\dagger}(\vy)\right\} &= \Lambda_{\pm}(\vx, \vy) \, , & \quad\quad \left\{\psi_{\pm}(\vx), \psi_{\mp}^{\dagger}(\vy)\right\} &= 0 \\
\Bigl\{\psi_{\pm}(\vx), \psi_{\pm}(\vy)\Bigr\} &= 0 \, , & \quad\quad \Bigl\{\psi_{\pm}(\vx), \psi_{\mp}(\vy)\Bigr\} &= 0 \, .
\end{alignat}
\label{Gl: Antikommutator1}%
\end{subequations}

In the following, we will consider only the limit of chiral quarks, i.e.~$m_{\mathrm{Q}} = 0$. The extension of our ansatz to massive quarks would be, however, straightforward. The quark trial vacuum state is chosen as the most general Slater determinant which is not orthogonal to the bare vacuum $\vert 0 \rangle$. By Thouless's theorem such a state can be expressed as
\beq
\vert \phi_{\mathrm{Q}}[A] \rangle = \exp\left[-\int \dd^3 x \int \dd^3 y \, \psi_+^{\dagger}(\vx) K(\vx, \vy) \psi_-(\vy)\right] \vert 0 \rangle \, , \label{Gl: VakuumfunktionalQ}
\eeq
where the (gauge field dependent) kernel $K$ is a matrix in the indices of the quark fields, i.e. in Lorentz and color space and, when different flavors are included, also in flavor space. The use of a Slater determinant allows the application of Wick's theorem, which facilitates the evaluation of the quark expectation value considerably. The norm of the fermionic wave functional (\ref{Gl: VakuumfunktionalQ}) is given by \cite{Pak2013}
\beq
I[A] \equiv \langle \phi_{\mathrm{Q}}[A] \vert \phi_{\mathrm{Q}}[A] \rangle = \det\begin{pmatrix}
                                                                                   \Id & K \\
                                                                                   K^{\dagger} & -\Id
                                                                                  \end{pmatrix}
= \det\bigl(\Id + K^{\dagger} K\bigr) \, , \label{Gl: FermiDeterminante1}
\eeq
where the first (functional) determinant is defined in the complete Hilbert space of the Dirac  Hamiltonian (\ref{Gl: DiracOperator}), while the second one is defined in the subspace of negative energy eigenfunctions only. Note that the fermion determinant $I[A]$ [eq.~(\ref{Gl: FermiDeterminante1})] explicitly depends on the gauge field $A$ and is therefore non-trivial. The kernel $K$ connects the positive with the negative energy subspace of the (single particle) Hilbert space of the Dirac Hamiltonian (\ref{Gl: DiracOperator}) and is chosen in the form
\beq
K (\vx, \vy) = \beta S(\vx, \vy) + g \int \dd^3 z \, \bigl[V(\vx, \vy; \vz) + \beta W(\vx, \vy; \vz)\bigr] \valpha \cdot \vA^a(\vz) t^a \, , \label{Gl: VakuumfunktionalQ1}
\eeq
where $S$, $V$ and $W$ are variational kernels, which, by translational invariance, depend only on the coordinate differences. For $V = W = 0$ our ansatz $\vert \phi_{\mathrm{Q}}[A] \rangle$ [eq.~(\ref{Gl: VakuumfunktionalQ})] reduces to the BCS-type wave functional considered in refs.~\cite{FM1982, Adler1984, AA1988}. For the BCS-wave functional the quark-gluon coupling term of the Dirac Hamiltonian, $H_{\mathrm{Q}}^A$ [eq.~(\ref{Gl: DiracHamiltonian})], escapes the expectation value. Such a wave functional does, however, already produce spontaneous breaking of chiral symmetry but not of sufficient amount. It yields, for $\sigma_{\mathrm{C}} = 2 \sigma$, a quark condensate $\langle \bar{\psi} \psi \rangle$ of about $(-165 \, \mathrm{MeV})^3$ which is significantly smaller than the phenomenological value of \cite{Williams2007}
\beq
\langle \bar{\psi}(\vx) \psi(\vx) \rangle_{\mathrm{phen}} = (-235 \, \mathrm{MeV})^3 \, . \label{Gl: ChiralesKondensat}
\eeq
For $W = 0$ the wave functional (\ref{Gl: VakuumfunktionalQ}) corresponds to the ansatz considered in refs.~\cite{Pak2012a, Pak2013}, where it was shown that the inclusion of the  explicit coupling of the quarks to the gluons by the term proportional to $V$ gives a substantial improvement compared to the BCS-type wave functional ($V = W = 0$). Here we go one step further and include also the coupling term proportional to $W$. As we will show below this term does not only improve the previous variational calculation because of the enlarged space of trial states but has the principle advantage that all linear ultraviolet divergences disappear from the gap equation for the scalar kernel $S$. Furthermore, the presence of this kernel can be motivated by perturbation theory on top of a BCS-vacuum state, see appendix \ref{Anh: Stoerungstheorie}.

It is convenient to include the fermion determinant $I$ [eq.~(\ref{Gl: FermiDeterminante1})] in our ansatz for the Yang--Mills part $\phi_{\mathrm{YM}}$ of the vacuum wave functional (\ref{Gl: Vakuumfunktional}) in the same way as the Faddeev--Popov determinant $J$. Therefore, we choose the following ansatz
\begin{subequations}
\begin{align}
\phi_{\mathrm{YM}}[A] &=  \Nc I^{-\frac{1}{2}}[A] J^{-\frac{1}{2}}[A] \widetilde{\phi}_{\mathrm{YM}}[A] \, , \\
\widetilde{\phi}_{\mathrm{YM}}[A] &= \exp\left(-\frac{1}{2} \int \dd^3 x \int \dd^3 y \, A_k^a(\vx) \omega(\vx, \vy) A_k^a(\vy)\right) \, ,
\end{align}
\label{Gl: BoseAnsatz1}%
\end{subequations}
which differs from (\ref{Gl: VakuumfunktionalYM}) only by the presence of the fermion determinant. Again, $\omega$ is the variational kernel to be determined by minimizing the ground state energy. In terms of the functional (\ref{Gl: BoseAnsatz1}) the expectation value of an arbitrary operator
$O$ [eq.~(\ref{Gl: EW})] reads
\beq
\langle O[A, \Pi, \psi] \rangle = |\Nc|^2 \int \Dc A \, \widetilde{\phi}^*_{\mathrm{YM}}[A] \langle \widetilde{O}[A, \Pi, \psi] \rangle_{\mathrm{Q}} \, \widetilde{\phi}_{\mathrm{YM}}[A] \label{Gl: EW1}
\eeq
where we have introduced the transformed operator
\beq
\widetilde{O}[A, \Pi, \psi] \equiv J^{\frac{1}{2}}[A] I^{\frac{1}{2}}[A] O[A, \Pi, \psi] J^{-\frac{1}{2}}[A] I^{-\frac{1}{2}}[A] \label{Gl: TransfOp}
\eeq
and the fermionic expectation value
\beq
\langle O \rangle_{\mathrm{Q}} = I^{-1} \langle \phi_{\mathrm{Q}} \vert O \vert \phi_{\mathrm{Q}} \rangle \, . \label{Gl: FermiEW1}
\eeq
For an operator $O$ which is independent of the canonical momentum $\Pi$, $O = \widetilde{O}$ holds and both the Faddeev--Popov and the fermion determinants disappear from the expectation value (\ref{Gl: EW1}) implying that Wick's theorem holds in the form
\beq
\langle O[A, \psi] \rangle = \left. \left\{\exp\left(\frac{1}{2} \int \dd^3 x \int \dd^3 y \, \frac{\delta}{\delta A_k^a(\vx)} D_{k l}^{a b}(\vx, \vy) \frac{\delta}{\delta A_l^b(\vy)}\right) \langle O[A, \psi]\rangle_{\mathrm{Q}}\right\}\right\vert_{A = 0} \,. \label{Gl: Wick1}
\eeq
If, however, the operator $O$ explicitly depends on the canonical momentum operator, the Faddeev--Popov determinant $J$ and the fermion determinant $I$ remain in the expectation value (\ref{Gl: EW1}), (\ref{Gl: TransfOp}) and the functional derivative imposed by the momentum operator $\Pi = \delta / \ii \delta A$ has to be carried out before Wick's theorem can be applied.

\section{Static quark propagator and chiral condensate} \label{Abschn: Propagator}

Since our quark wave functional is a Gaussian in the fermion fields Wick's theorem holds and as a consequence pure fermionic expectation values can be expressed in terms of the static quark propagator
\beq
G_{i j}^{m n}(\vx, \vy) = \frac{1}{2} \langle \Bigl[\psi_i^m(\vx), {\psi_j^n}^{\dagger}(\vy)\Bigr] \rangle \, . \label{Gl: StatProp}
\eeq
For the fermionic expectation value of the quark bilinear in the state $\vert \phi_{\mathrm{Q}}[A] \rangle$ [eq.~(\ref{Gl: VakuumfunktionalQ})] one finds \cite{Pak2013}
\begin{subequations}
\begin{align}
\langle \psi_{+, i}(\vx) \psi_{+, j}^{\dagger}(\vy) \rangle_{\mathrm{Q}} &= \bigl(\Lambda_+ \bigl[\Id + K K^{\dagger}\bigr]^{-1} \Lambda_+\bigr)_{i j}(\vx, \vy) \, , \\
\langle \psi_{-, i}^{\dagger}(\vx) \psi_{-, j}(\vy) \rangle_{\mathrm{Q}} &= \bigl(\Lambda_- \bigl[\Id + K^{\dagger} K\bigr]^{-1} \Lambda_-)_{j i}(\vy, \vx) \, , \\
\langle \psi_{-, i}(\vx) \psi_{+, j}^{\dagger}(\vy) \rangle_{\mathrm{Q}} &= \bigl(\Lambda_- \bigl[\Id + K^{\dagger} K]^{-1} K^{\dagger} \Lambda_+\bigr)_{i j}(\vx, \vy) \, , \\
\langle \psi_{+, i}(\vx) \psi_{-, j}^{\dagger}(\vy) \rangle_{\mathrm{Q}} &= \bigl(\Lambda_+ \bigl[\Id + K K^{\dagger}]^{-1} K \Lambda_-)_{i j}(\vx, \vy) \, .
\end{align}
\label{Gl: Operatorpaerchen}%
\end{subequations}
Unfortunately, the evaluation of the bosonic expectation value is a bit more involved and cannot be carried out without further approximations. This is because the fermionic two-point functions always contain a term like
\beq
\left[\Id + K K^{\dagger}\right]^{-1} = \sum_{n = 0}^{\infty} (-1)^n \left(K K^{\dagger}\right)^n
\eeq
which is an infinite series in the gauge field $A$ so that its bosonic expectation value cannot be evaluated in closed form. For calculating the ground state energy up to two-loop order, it is sufficient to expand the exponential in (\ref{Gl: Wick1}) in a Taylor series up to leading order yielding
\beq
\langle O[A, \psi] \rangle \approx \langle O[A = 0, \psi]\rangle_{\mathrm{Q}} + \frac{1}{2} \int \dd^3 x \int \dd^3 y \, \frac{\delta}{\delta A_k^a(\vx)} D_{k l}^{a b}(\vx, \vy) \frac{\delta}{\delta A_l^b(\vy)} \langle O[A, \psi]\rangle_{\mathrm{Q}}\Bigr\vert_{A = 0}. \label{Gl: Wick2}
\eeq
For the static quark propagator
\beq
G_{i j}^{m n}(\vx, \vy) = \delta^{m n} \int \da^3 p \, \exp\bigl(\ii \vp \cdot (\vx - \vy)\bigr) G_{i j}(\vp)
\eeq
in momentum space we find then after somewhat lengthy calculations, see appendix \ref{Anh: Propagator},
\beq
G(\vp) = \frac{P(p)}{2} \Bigl[1 - S^2(p) - I_{\alpha}(p)\Bigr] \valpha \cdot \hp + P(p) \Bigl[S(p) - I_{\beta}(p)\Bigr] \beta \label{Gl: StatProp1}
\eeq
where $\hp = \vp / p$ and
\begin{align}
I_{\alpha}(p) &= C_{\mathrm{F}} g^2 \int \da^3 q \, \frac{P(p) P(q)}{\omega(|\vp + \vq|)} \left[V^2(\vp, \vq) X(\vp, \vq) \Bigl(1 + 2 S(p) S(q) - S^2(p)\Bigr) \right. \nonumber \\
&\phantom{=}\,\, \phantom{C_{\mathrm{F}} g^2 \int \da^3 q \, \frac{P(p) P(q)}{\omega(|\vp + \vq|)} \left[\right.} \left.+ W^2(\vp, \vq) Y(\vp, \vq) \Bigl(1 - 2 S(p) S(q) - S^2(p)\Bigr)\right] \label{Gl: Schleifenintegralalpha} \\
I_{\beta}(p) &= \frac{C_{\mathrm{F}}}{2} g^2 \int \da^3 q \, \frac{P(p) P(q)}{\omega(|\vp + \vq|)} \left[V^2(\vp, \vq) X(\vp, \vq) \Bigl(2 S(p) - S(q) + S^2(p) S(q)\Bigr) \right. \nonumber \\
&\phantom{=}\,\, \phantom{\frac{C_{\mathrm{F}}}{2} g^2 \int \da^3 q \, \frac{P(p) P(q)}{\omega(|\vp + \vq|)} \left[\right.} \left. + W^2(\vp, \vq) Y(\vp, \vq) \Bigl(2 S(p) + S(q) - S^2(p) S(q)\Bigr)\right] . \label{Gl: Schleifenintegralbeta}
\end{align}
Here we have replaced the variational kernels by their respective momentum space representation, see appendix \ref{Anh: Impulsdarstellung}, and, furthermore, introduced the abbreviations
\begin{align}
P(p) &= \frac{1}{1 + S^2(p)} \label{Gl: PFaktor} \\
X(\vp, \vq) &= 1 - \Bigl[\hp \cdot \hpq\Bigr] \Bigl[\hq \cdot \hpq\Bigr] \\
Y(\vp, \vq) &= 1 + \Bigl[\hp \cdot \hpq\Bigr] \Bigl[\hq \cdot \hpq\Bigr]
\end{align}
as well as the Casimir factor $C_{\mathrm{F}} = (\Ncc^2 - 1) / 2 \Ncc$. Neglecting the coupling of the quarks to the transversal gluons, $V = W = 0$, the propagator (\ref{Gl: StatProp1}) reduces to the BCS result obtained in ref.~\cite{Adler1984}. However, even when we include the coupling term $\sim V$ but ignore the additional coupling term $\sim W$ our quark propagator differs from that of ref.~\cite{Pak2013}, although in that case
the trial ans\"atze for the quark wave functional agree. The reason is that in ref.~\cite{Pak2013} a simplifying approximation was used in the evaluation of the gluonic expectation value of fermionic objects: In the denominators of the quark propagator (\ref{Gl: Operatorpaerchen}) the kernel $K^{\dagger} K$ was replaced by its (gluonic) vacuum expectation value $\langle K^{\dagger} K \rangle$. In the present paper we abandoned this approximation and strictly carried out the calculation to two-loop order. The results obtained in this way are also consistent with those obtained in the Dyson--Schwinger approach \cite{CR201Xa} and in perturbation theory, as we will see later.

Due to the fact that our quark wave functional is a Slater determinant the static quark propagator (\ref{Gl: StatProp1}) can be brought to the quasi-particle form
\beq
G(\vp) = Z(p) \frac{\valpha \cdot \vp + \beta M(\vp)}{2 \sqrt{p^2 + M^2(p)}} \label{Gl: StatProp3}
\eeq
with an effective (running) mass
\beq
M(p) = \frac{2 p \bigl[S(p) - I_{\beta}(p)\bigr]}{1 - S^2(p) - I_{\alpha}(p)} \label{Gl: Massenfkt}
\eeq
and a field renormalization factor
\beq
Z(p) = P(p) \sqrt{\bigl[1 - S^2(p) - I_{\alpha}(p)\bigr]^2 + 4 \bigl[S(p) - I_{\beta}(p)\bigr]^2} \,. \label{Gl: Feldrenormierung}
\eeq
If one neglects the coupling of the quarks to the gluons the two loop integrals vanish, $I_{\alpha} = I_{\beta} = 0$, and we recover the result of ref.~\cite{Adler1984}
\beq
M(p) = \frac{2 p S(p)}{1 - S^2(p)}, \quad\quad Z(p) = 1 \,. \label{Gl: ADMasse}
\eeq
From this expression it is already clear that a non-vanishing scalar kernel $S$ implies a non-zero effective quark mass function and thus spontaneous breaking of chiral symmetry. Indeed the order parameter of spontaneous breaking of chiral symmetry, the quark condensate, can be expressed by means of the static quark propagator (\ref{Gl: StatProp}) as
\beq
\langle \bar{\psi}(\vx) \psi(\vx) \rangle = - \tr\bigl(\beta G(\vx, \vx)\bigr) \, . \label{Gl: ChiralesKondensat3}
\eeq
Inserting here the explicit form of the propagator (\ref{Gl: StatProp3}) we find
\beq
\langle \bar{\psi}(\vx) \psi(\vx) \rangle = -2 \int \da^3 p \, \frac{Z(\vp) M(\vp)}{\sqrt{p^2 + M^2(\vp)}} \, . \label{Gl: ChiralesKondensat4}
\eeq
Obviously, the quark condensate vanishes when no effective mass is generated, i.e. for $M = 0$.

\section{Ground state energy} \label{Abschn: EW}

Below we evaluate the expectation value of the QCD Hamiltonian (\ref{Gl: QCDHamiltonian1}) in our trial state (\ref{Gl: Vakuumfunktional}) [with eqs.~(\ref{Gl: VakuumfunktionalQ}) and (\ref{Gl: BoseAnsatz1})]. The calculations will be carried out consistently to two-loop level.

Since the quark wave functional depends explicitly on the gauge field it is mandatory to take the fermionic expectation value before the bosonic one. We begin with the expectation value of the Dirac Hamiltonian $H_{\mathrm{Q}}$.

\subsection{Quark energy}

\begin{figure}
\centering
\includegraphics[width=0.35\linewidth]{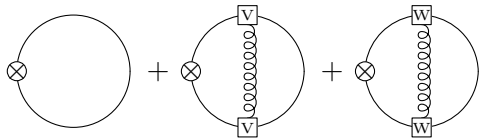}
\caption{Diagrammatic representation of the expectation value of the free Dirac Hamiltonian (\ref{Gl: FrDiracHamEW}). 
{We denote the free Dirac operator by a crossed circle and the vector kernels $V$ and $W$ by a labeled square. Straight and curly lines stand, respectively, for the quark and gluon propagator.}}
\label{Abb: FeynmanFrDirac}
\end{figure}

The Dirac Hamiltonian (\ref{Gl: DiracHamiltonian}) consists of two parts, one describing a free Dirac particle and the other containing the coupling between quarks and transversal gluons. The expectation value of the free Dirac Hamiltonian can be expressed by the quark Green function
\beq
\langle H_{\mathrm{Q}}^0 \rangle = -\Ncc \deltabar^3(0) \int \da^3 p \, \tr\Bigl(\valpha \cdot \vp \, G(\vp)\Bigr) \, , \label{Gl: FrDiracHamEW2}
\eeq
where $\deltabar^3(0) = \int \dd^3 x$ is the (infinite) spatial volume ($\deltabar = 2 \pi \delta$). With the explicit form (\ref{Gl: StatProp3}) we find
\beq
\langle H_{\mathrm{Q}}^0 \rangle = -2 \Ncc \deltabar^3(0) \int \da^3 p \, \frac{p^2 Z(p)}{\sqrt{p^2 + M^2(p)}} \,. \label{Gl: FrDiracHamEW}
\eeq
Here $G$, $M$ and $Z$ are functionals of the variational kernels $S$, $V$ and $W$ [see eqs.~(\ref{Gl: StatProp3}), (\ref{Gl: Massenfkt}) and (\ref{Gl: Feldrenormierung})]. Using the explicit expression for these quantities one finds for $\langle H_{\mathrm{Q}}^0 \rangle$ a somewhat lengthy expression, which is given in eq.~(\ref{Gl: FrDiracHamEW1}) of appendix \ref{Anh: EW}. The obtained expression for $\langle H_{\mathrm{Q}}^0 \rangle$ allows, however, for a direct interpretation in terms of Feynman diagrams given in fig.~\ref{Abb: FeynmanFrDirac}.

Let us also mention that there is no need to subtract the energy of the trivial vacuum, $\langle H_{\mathrm{QCD}} \rangle\vert_{S = V = W = 0}$, because this would only shift the vacuum energy by an irrelevant constant.

\begin{figure}
\centering
\includegraphics[width=0.25\linewidth]{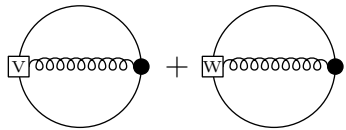}
\caption{Diagrammatic representation of the expectation value of the quark-gluon coupling, $\langle H_{\mathrm{Q}}^A \rangle$ (\ref{Gl: Kopplungsterm}). The filled dot stands for the bare quark-gluon vertex in the Hamiltonian $H_{\mathrm{Q}}^A$.}
\label{Abb: FeynmanKopplungsterm}
\end{figure}

The evaluation of the coupling term $\langle H_{\mathrm{Q}}^A \rangle$ is more involved due to the presence of the gauge field in $H_{\mathrm{Q}}^A$ but can be straighforwardly carried out to two loops. One finds then the following expression
\begin{align}
\langle H_{\mathrm{Q}}^A \rangle &= -\bigl(\Ncc^2 - 1\bigr) \deltabar^3(0) g^2 \int \da^3 p \int \da^3 q \, \frac{V(\vp, \vq)}{\omega(|\vp + \vq|)} P(q) P(p) \Bigl(1 + S(q) S(p)\Bigr) X(\vp, \vq) \nonumber \\
&\phantom{=}\,\, -\bigl(\Ncc^2 - 1\bigr) \deltabar^3(0) g^2 \int \da^3 p \int \da^3 q \, \frac{W(\vp, \vq)}{\omega(|\vp + \vq|)} P(q) P(p) \Bigl(S(p) + S(q)\Bigr) Y(\vp, \vq) \, , \label{Gl: Kopplungsterm}
\end{align}
which is diagrammatically illustrated in fig.~\ref{Abb: FeynmanKopplungsterm}.

\begin{figure}
\centering
\includegraphics[width=0.1\linewidth]{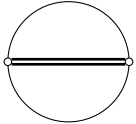}
\caption{Diagrammatic representation of the expectation value of the fermionic part of the color Coulomb potential (\ref{Gl: CoulombQQ}). The  double line stands for the Coulomb kernel [eq.~$V_{\mathrm{C}}$ (\ref{Gl: Coulombkern3})].}
\label{Abb: FeynmanCoulomb}
\end{figure}

The quark contribution to the Coulomb energy $\langle H_{\mathrm{C}}^{\mathrm{Q}} \rangle$ can be straighforwardly evaluated by using Wick's theorem in the quark sector. Due to the replacement of the Coulomb kernel $\hat{F}$ [eq.~(\ref{Gl: Coulombkern})] by its expectation value $V_{\mathrm{C}}$ [eq.~(\ref{Gl: Coulombkern3})], which is correct to two-loop order, we are left here with a quark two-body operator $\psi^{\dagger} \psi \, \psi^{\dagger} \psi$, whose fermionic expectation value leads to terms of the form $\langle \psi^{\dagger} \psi \rangle_{\mathrm{Q}} \langle \psi \, \psi^{\dagger} \rangle_{\mathrm{Q}}$. Furthermore, up to two loops the remaining gluonic expectation value can be taken for each fermion contraction $\langle \psi \, \psi^{\dagger} \rangle_{\mathrm{Q}}$ separately, which produces a static quark propagator (\ref{Gl: StatProp}). Exploiting the fact that the quark propagator is a color singlet and that $\tr \, t^a = 0$ one arrives at the following result
\beq
\langle H_{\mathrm{C}}^{\mathrm{Q}} \rangle = \frac{1}{8} t_{m_1 m_2}^a t_{m_3 m_4}^a \int \dd^3 x \int \dd^3 y \, V_{\mathrm{C}}(|\vx - \vy|) \Bigl[\delta^{m_1 m_4} \delta^{m_2 m_3} \delta_{i i} \delta^3(0) \delta^3(\vx - \vy) - 4 \mathrm{tr}\bigl(G^{m_4 m_1}(\vy, \vx) G^{m_2 m_3}(\vx, \vy)\bigr)\Bigr] \,.
\eeq
Inserting here the explicit expression for the static quark propagator (\ref{Gl: StatProp1}) and confining oneself to two-loop terms yields
\beq
\langle H_{\mathrm{C}}^{\mathrm{Q}} \rangle = \frac{\Ncc^2 - 1}{4} \deltabar^3(0) \int \da^3 p \int \da^3 q \, V_{\mathrm{C}}(|\vp - \vq|) \Bigl[1 - P(p) P(q) \Bigl(4 S(p) S(q) + \bigl(1 - S^2(p)\bigr) \bigl(1 - S^2(q)\bigr) \hp \cdot \hq\Bigr)\Bigr] \, . \label{Gl: CoulombQQ}
\eeq
The same result is found for the BCS-wave functional ($V = W = 0$). This is because the quark-gluon coupling vertices in our trial wave functional (\ref{Gl: VakuumfunktionalQ}), (\ref{Gl: VakuumfunktionalQ1}) contribute only loop terms to the static quark propagator, see eq.~(\ref{Gl: StatProp1}), which, when kept, would produce three-loop terms in $\langle H_{\mathrm{C}}^{\mathrm{Q}} \rangle$. The quark contribution to the Coulomb energy (\ref{Gl: CoulombQQ}) is diagrammatically illustrated in fig.~\ref{Abb: FeynmanCoulomb}.

\subsection{Energy of transversal gluons}

\begin{figure}
\centering
\includegraphics[width=0.25\linewidth]{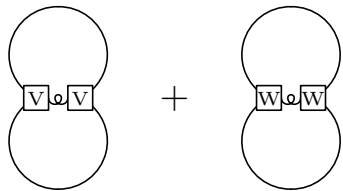}
\caption{Diagrammatic representation of the fermionic contributions to the expectation value of the kinetic energy of the transversal gluons (\ref{Gl: Kinen1}).}
\label{Abb: FeynmanKinEn}
\end{figure}

We continue with the contribution of the kinetic energy of the transversal gluons $H_{\mathrm{YM}}^E$, see eq.~(\ref{Gl: YMHamiltonian}). Although the operator $H_{\mathrm{YM}}^E$ does not contain the quark field, the quarks do contribute to its expectation value due to the action of the momentum operator on the gauge field in the quark wave functional, see eqs.~(\ref{Gl: VakuumfunktionalQ}) and (\ref{Gl: VakuumfunktionalQ1}). Furthermore, the momentum operator does not commute with the fermion determinant (\ref{Gl: FermiDeterminante1}). As a consequence the transformed operator $\widetilde{H}_{\mathrm{YM}}^E$ [eq.~(\ref{Gl: TransfOp})] becomes non-trivial\footnote{Notice that canonical momentum operators inside of square brackets do not act on terms which stand outside of the respective bracket.}
\begin{align}
\widetilde{H}_{\mathrm{YM}}^E &= \frac{1}{2} \int \da^3 p \left(\Pi_k^a(\vp) \Pi_k^a(-\vp) - \bigl[\Pi_k^a(\vp) \ln I \bigr] \Pi_k^a(-\vp) - \frac{1}{4} \bigl[\Pi_k^a(\vp) \ln J \bigr] \bigl[\Pi_k^a(-\vp) \ln J \bigr] \right. \nonumber \\
&\phantom{=}\,\, \quad\quad + \left. \frac{1}{4} \bigl[\Pi_k^a(\vp) \ln I \bigr] \bigl[\Pi_k^a(-\vp) \ln I \bigr] - \frac{1}{2} \bigl[\Pi_k^a(\vp) \Pi_k^a(-\vp) \ln J \bigr] - \frac{1}{2} \bigl[\Pi_k^a(\vp) \Pi_k^a(-\vp) \ln I \bigr]\right). \label{Gl: TransfKinEn}
\end{align}
Obviously this transformed operator has a much more complicated structure than the original one. This is the price we have to pay for the elimination of the Faddeev--Popov and quark determinants from the gluonic integration measure of the scalar product, see eq.~(\ref{Gl: EW1}). The evaluation of the expectation value of $\widetilde{H}_{\mathrm{YM}}^E$ [eq.~(\ref{Gl: TransfKinEn})] is quite involved and sketched in appendix \ref{Anh: EW}. Up to two-loop order one finds the following expression:
\begin{align}
\langle H_{\mathrm{YM}}^E \rangle &= \frac{\Ncc^2 - 1}{2} \deltabar^3(0) \int \da^3 p \, \frac{\bigl(\omega(p) - \chi(p)\bigr)^2}{\omega(p)} \nonumber \\
&\phantom{=}\,\, + \frac{\Ncc^2 - 1}{2} \deltabar^3(0) g^2 \int \da^3 p \int \da^3 q \, P(p) P(q) V^2(\vp, \vq) X(\vp, \vq) \nonumber \\
&\phantom{=}\,\, + \frac{\Ncc^2 - 1}{2} \deltabar^3(0) g^2 \int \da^3 p \int \da^3 q \, P(p) P(q) W^2(\vp, \vq) Y(\vp, \vq) \label{Gl: Kinen1}
\end{align}
Here $\chi(p) = \delta^{a b} \tr \, t(\vp) \chi^{a b}(\vp) / (2 (\Ncc^2 - 1))$ is the scalar curvature. The first term in eq.~(\ref{Gl: Kinen1}) arises from the Yang--Mills part of the vacuum wave functional and was already obtained in ref.~\cite{Feuchter2004, Feuchter2004a}. The last two terms give the quark contributions which are diagrammatically represented in  fig.~\ref{Abb: FeynmanKinEn}.

The potential energy of the transversal gluons $H_{\mathrm{YM}}^B$ [eq.~(\ref{Gl: YMHamiltonian})] is a functional of the gauge field $A$ only. As a consequence the quarks do not contribute to $\langle H_{\mathrm{YM}}^B \rangle$ which is hence  given by the expresson obtained in ref.~\cite{Feuchter2004, Feuchter2004a} for the Yang--Mills sector
\beq
\langle {H}_{\mathrm{YM}}^B \rangle = \frac{\Ncc^2 - 1}{2} \deltabar^3(0) \int \da^3 p \, \frac{p^2}{\omega(p)} + \frac{\Ncc \left(\Ncc^2 - 1\right)}{16} \deltabar^3(0) g^2 \int \da^3 p \int \da^3 q \, \frac{3 - \left(\hp \cdot \hq\right)^2}{\omega(p) \omega(q)} \, .\label{Gl: PotEnergie}
\eeq

Finally, we calculate the expectation value of the purely gluonic part of the Coulomb term, $H_{\mathrm{C}}^{\mathrm{YM}}$. Although this operator contains the momentum operator $\Pi$ the quarks do not contribute to two-loop order and we obtain the same result as in pure Yang--Mills theory \cite{Feuchter2004, Feuchter2004a}
\beq
\langle H_{\mathrm{C}}^{\mathrm{YM}} \rangle = \frac{N_{\mathrm{C}} \left(N_{\mathrm{C}}^2 - 1\right)}{16} \deltabar^3(0) \int \da^3 p \int \da^3 q \, V_{\mathrm{C}}(|\vp - \vq|) \frac{\bigl(\omega(p) - \chi(p) - \omega(q) + \chi(q)\bigr)^2}{\omega(p) \omega(q)} \left(1 + \left(\hp \cdot \hq\right)^2\right) \,. \label{Gl: CoulombYM}
\eeq

\subsection{Total energy}

As already mentioned before, the mixed Coulomb term $H_{\mathrm{C}}^{\mathrm{INT}}$ does not contribute to two-loop order. The total vacuum energy is thus given by
\beq
\langle H_{\mathrm{QCD}} \rangle =  \langle \bar{H}_{\mathrm{YM}} \rangle + \langle \bar{H}_{\mathrm{Q}} \rangle \, , \label{Gl: EnergieEW}
\eeq
where the various contributions to the gluon energy 
\beq
\langle \bar{H}_{\mathrm{YM}} \rangle = \langle H_{\mathrm{YM}}^E \rangle + \langle H_{\mathrm{YM}}^B \rangle + \langle H_{\mathrm{C}}^{\mathrm{YM}} \rangle
\eeq
are given by eqs.~(\ref{Gl: Kinen1}), (\ref{Gl: PotEnergie}) and (\ref{Gl: CoulombYM}), while the contribution to the energy of the quarks interacting with the gluons
\beq
\langle \bar{H}_{\mathrm{Q}} \rangle = \langle H_{\mathrm{Q}}^0 \rangle + \langle H_{\mathrm{Q}}^A \rangle + \langle H_{\mathrm{C}}^{\mathrm{Q}} \rangle
\eeq
are given by eqs.~(\ref{Gl: FrDiracHamEW}) (see also eq.~(\ref{Gl: FrDiracHamEW1})), (\ref{Gl: Kopplungsterm}) and (\ref{Gl: CoulombQQ}). At this point it is  worth to compare the present result with that of previous work. If one neglects the coupling between quarks and transversal gluons, $V = W = 0$, 
$\langle \bar{H}_{\mathrm{YM}} \rangle$ becomes the vacuum energy of the Yang--Mills sector obtained in ref.~\cite{Feuchter2004, Feuchter2004a} and $\langle \bar{H}_{\mathrm{Q}} \rangle$ reduces to the vacuum energy of the model considered in ref.~\cite{Adler1984}. When the quark-gluon coupling term $\sim V$ is included but the other coupling term ignored, $W = 0$, one recovers the result of the Dyson--Schwinger approach of ref.~\cite{CR201Xa}, which differs, however, from the result of ref.~\cite{Pak2013} due to further simplifying approximations used there, see above. Finally, when both quark-gluon coupling terms are included in the quark wave functional but the trivial solution of the gap equation $S = 0$ is assumed one recovers from $\langle \bar{H}_{\mathrm{Q}} \rangle$ the quark energy in second order perturbation theory \cite{CR2015a}.

\section{Variational equations} \label{Abschn: Variation}

\begin{figure}%
\centering%
\parbox{0.125\linewidth}{%
\centering%
\includegraphics[width=\linewidth]{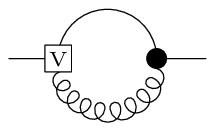} \\%
(a) %
}%
\hspace{0.1\linewidth}%
\parbox{0.125\linewidth}{%
\centering%
\includegraphics[width=\linewidth]{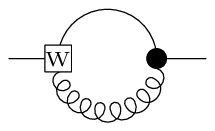} \\%
(b) %
}%
\caption{Diagrammatic representation of the contribution of the quark-gluon coupling in the Dirac Hamiltonian to the quark gap equation, (a) eq.~(\ref{Gl: KopplungstermGapglV}) and (b) eq.~(\ref{Gl: KopplungstermGapglW}).}%
\label{Abb: FeynmanKopplungstermGapgl}%
\end{figure}

\begin{figure}%
\centering%
\parbox{0.125\linewidth}{%
\centering%
\includegraphics[width=\linewidth]{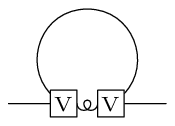} \\%
(a) %
}%
\hspace{0.1\linewidth}%
\parbox{0.125\linewidth}{%
\centering%
\includegraphics[width=\linewidth]{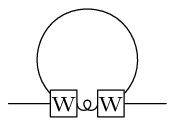} \\%
(b) %
}%
\caption{Diagrammatic representation of the contribution of the kinetic energy of the transversal gluons to the quark gap equation. (a) and (b) correspond, respectively, to the first and second term on the r.h.s. of eq.~(\ref{Gl: KinetischeEnergieGapgl}).}%
\label{Abb: FeynmanKinEnGapgl}%
\end{figure}

Our ansatz for the vacuum functional [(\ref{Gl: Vakuumfunktional}) with (\ref{Gl: VakuumfunktionalQ}) and (\ref{Gl: BoseAnsatz1})] contains four variational kernels $\omega$, $S$, $V$ and $W$, which we will determine in the following according to the variational principle. From $\delta \langle H_{\mathrm{QCD}} \rangle / \delta S(k) = 0$ we find the following integral equation
\beq
k S(k) = I_{\mathrm{C}}^{\mathrm{Q}}(k) + I_{V V}^{\mathrm{Q}}(k) + I_{W W}^{\mathrm{Q}}(k) + I_{V \mathrm{Q}}^{\mathrm{Q}}(k) + I_{W \mathrm{Q}}^{\mathrm{Q}}(k) + I_{E}^{\mathrm{Q}}(k) \, , \label{Gl: Gapgleichung}
\eeq
to which we will refer as quark gap equation. Here
\beq
I_{\mathrm{C}}^{\mathrm{Q}}(k) = \frac{C_{\mathrm{F}}}{2} \int \da^3 p \, V_{\mathrm{C}}(|\vp - \vk|) P(p) \left[S(p) \bigl(1 - S^2(k)\bigr) - S(k) \bigl(1 - S^2(p)\bigr) \hp \cdot \hk\right] \label{Gl: CoulombQQGapgl}
\eeq
\begin{figure}%
\centering%
\includegraphics[width=0.15\linewidth]{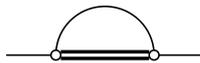}%
\caption{Diagrammatic representation of the contribution (\ref{Gl: CoulombQQGapgl}) of the Coulomb potential to the quark gap equation.}%
\label{Abb: FeynmanCoulombGapgl}%
\end{figure}%
is the contribution from the Coulomb term $H_{\mathrm{C}}^{\mathrm{Q}}$ which is illustrated in fig.~\ref{Abb: FeynmanCoulombGapgl}. Only this term survives when the coupling of the quarks to the transversal gluons in the vacuum wave functional is switched off ($V = W = 0$). Furthermore,
\begin{align}
I_{V V}^{\mathrm{Q}}(k) &= -\frac{C_{\mathrm{F}}}{2} g^2 \int \da^3 p \, \frac{V^2(\vp, \vk)}{\omega(|\vp + \vk|)} X(\vp, \vk) P(p) \Bigl\{k P(k) S(k)\Bigl[-3 + S^2(k)\Bigr] + p P(p) S(k) \Bigl[-1 + S^2(p)\Bigr] \nonumber \\
&\phantom{=}\,\, \phantom{-\frac{C_{\mathrm{F}}}{2} g^2 \int \da^3 p \, \times \Bigl\{} + k P(k) S(p) \Bigl[1 - 3 S^2(k)\Bigr] + p P(p) S(p) \Bigl[1 - S^2(k)\Bigr]\Bigr\} \label{Gl: FrDiracGapglV} \\
\intertext{and}
I_{W W}^{\mathrm{Q}}(k) &= - \frac{C_{\mathrm{F}}}{2} g^2 \int \da^3 p \, \frac{W^2(\vp, \vk)}{\omega(|\vp + \vk|)} Y(\vp, \vk) P(p) \Bigl\{k P(k) S(k) \Bigl[-3 + S^2(k)\Bigr] + p P(p) S(k) \Bigl[-1 + S^2(p)\Bigr] \nonumber \\
&\phantom{=}\,\, \phantom{- \frac{C_{\mathrm{F}}}{2} g^2 \int \da^3 p \, \times \Bigl\{} - k P(k) S(p) \Bigl[1 - 3 S^2(k)\Bigr] - p P(p) S(p) \Bigl[1 - S^2(k)\Bigr]\Bigr\} \label{Gl: FrDiracGapglW}
\end{align}
\begin{figure}%
\centering%
\parbox{0.125\linewidth}{%
\centering%
\includegraphics[width=\linewidth]{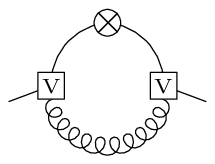} \\%
(a) %
}%
\hspace{0.1\linewidth}%
\parbox{0.125\linewidth}{%
\centering%
\includegraphics[width=\linewidth]{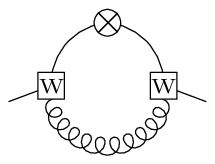} \\%
(b) %
}%
\caption{Diagrammatic representation of the contribution of the free Dirac Hamiltonian to the quark gap equation, (a) eq.~(\ref{Gl: FrDiracGapglV}) and (b) eq.~(\ref{Gl: FrDiracGapglW}).}%
\label{Abb: FeynmanFrDiracGapgl}%
\end{figure}%
result from the two-loop contribution of the free Dirac operator to the vacuum energy. These terms are graphically illustrated in fig.~\ref{Abb: FeynmanFrDiracGapgl}. The quark-gluon coupling in the Dirac Hamiltonian gives rise to the two diagrams shown in fig.~\ref{Abb: FeynmanKopplungstermGapgl} with either a $V$- or a $W$-vertex. Their contributions to the quark gap equation (\ref{Gl: Gapgleichung}) read
\begin{align}
I_{V \mathrm{Q}}^{\mathrm{Q}}(k) &= \frac{C_{\mathrm{F}}}{2} g^2 \int \da^3 p \, \frac{V(\vp, \vk)}{\omega(|\vp + \vk|)} X(\vp, \vk) P(p) \Bigl[S(p) \bigl(1 - S^2(k)\bigr) - 2 S(k)\Bigr] \label{Gl: KopplungstermGapglV} \\
I_{W \mathrm{Q}}^{\mathrm{Q}}(k) &= \frac{C_{\mathrm{F}}}{2} g^2 \int \da^3 p \, \frac{W(\vp, \vk)}{\omega(|\vp + \vk|)} Y(\vp, \vk) P(p) \Bigl[1 - S^2(k) - 2 S(k) S(p)\Bigr] \,. \label{Gl: KopplungstermGapglW}
\end{align}
Finally,
\beq
I_{E}^{\mathrm{Q}}(k) = \frac{C_{\mathrm{F}}}{2} g^2 S(k) \int \da^3 p \, V^2(\vp, \vk) X(\vp, \vk) P(p) + \frac{C_{\mathrm{F}}}{2} g^2 S(k) \int \da^3 p \, W^2(\vp, \vk) Y(\vp, \vk) P(p) \label{Gl: KinetischeEnergieGapgl}
\eeq
arises from the quark contribution to the kinetic energy of the transversal gluons [see the last two terms on the r.h.s. of eq.~(\ref{Gl: Kinen1})] and is illustrated in fig.~\ref{Abb: FeynmanKinEnGapgl}.

Neglecting the coupling between quarks and transversal gluons, $V = W = 0$, the quark gap equation (\ref{Gl: Gapgleichung}) reduces to the one obtained with a BCS-ansatz for the vacuum wave functional \cite{Adler1984}.

The variation  with respect to the vector kernel $V$, $\delta \langle H_{\mathrm{QCD}} \rangle / \delta V(\vk, \vk') = 0$, leads to an equation which can be explicitly solved for $V(\vk, \vk')$ yielding
\beq
V(\vk, \vk') = \frac{1 + S(k) S(k')}{k P(k) \Bigl(1 - S^2(k) + 2 S(k) S(k')\Bigr) + k' P(k') \Bigl(1 - S^2(k') + 2 S(k) S(k')\Bigr) + \omega(|\vk + \vk'|)} \,. \label{Gl: VKern}
\eeq
This result differs drastically from the one obtained in ref.~\cite{Pak2013}, where an integral equation for $V$ was obtained which could not be explictly solved. Furthermore, the vector kernel obtained there depends only on a single momentum argument. However, the quark-gluon vertex connecting three fields should, after taking into account overall momentum conservation, depend on two (independent) momentum arguments, as the vertex (\ref{Gl: VKern}) does. Finally, for the trivial solution of the gap equation, $S = 0$, the expression (\ref{Gl: VKern}) reduces to the perturbative result \cite{CR2015a}
\beq
V_0(\vk, \vk') = \frac{1}{k + k' + \omega(|\vk + \vk'|)} \,. \label{Gl: VKernStoerungstheorie}
\eeq

Minimization of the energy density with respect to the second vector kernel $W$ yields an equation which can, again, be solved directly:
\beq
W(\vk, \vk') = \frac{S(k) + S(k')}{k P(k) \Bigl(1 - S^2(k) - 2 S(k) S(k')\Bigr) + k' P(k') \Bigl(1 - S^2(k') - 2 S(k) S(k')\Bigr) + \omega(|\vk + \vk')|} \,. \label{Gl: WKern}
\eeq
As expected, the kernel $W$ also depends on two momenta. Although the structure of this equation is similar to the one for the kernel $V$ [eq.~(\ref{Gl: VKern})] there is an essential difference: The kernel $W$ [eq.~(\ref{Gl: WKern})] vanishes in the chirally symmetric phase $S = 0$ and is therefore only non-perturbatively realized, while the other vector kernel $V$ [eq.~(\ref{Gl: VKern})] reduces for $S = 0$ to the perturbative expression (\ref{Gl: VKernStoerungstheorie}).

Finally, for the bosonic kernel $\omega$ we obtain from $\delta \langle H_{\mathrm{QCD}} \rangle / \delta \omega(k) = 0$ the following integral equation
\beq
\omega^2(k) = \omega_{\mathrm{YM}}^2(k) + I_{V V}^{\mathrm{YM}}(k) + I_{W W}^{\mathrm{YM}}(k) + I_{V \mathrm{Q}}^{\mathrm{YM}}(k) + I_{W \mathrm{Q}}^{\mathrm{YM}}(k) \label{Gl: Gapglomega}
\eeq
where
\beq
\omega_{\mathrm{YM}}^2(k) = k^2 + \chi^2(k) + I_{\mathrm{T}}^{\mathrm{YM}} + I_{\mathrm{C}}^{\mathrm{YM}}(k) \label{Gl: GapglomegaYM}
\eeq
is the contribution from the gluonic energy $\langle \bar{H}_{\mathrm{YM}} \rangle$ with
\beq
I_{\mathrm{T}}^{\mathrm{YM}} = \frac{2 \Ncc}{3} g^2 \int \da^3 p \, \frac{1}{\omega(p)}
\eeq
being the gluonic tadpole and
\beq
I_{\mathrm{C}}^{\mathrm{YM}}(k) = \frac{\Ncc}{4} \int \da^3 p \, V_{\mathrm{C}}(|\vp - \vk|) \Bigl(1 + \bigl(\hp \cdot \hk\bigr)^2\Bigr) \frac{\bigl(\omega(p) - \chi(p) + \chi(k)\bigr)^2 - \omega^2(k)}{\omega(p)}
\eeq
being the contribution from the gluonic Coulomb term $\langle H_{\mathrm{C}}^{\mathrm{YM}} \rangle$. Equation (\ref{Gl: GapglomegaYM}) is obtained as gluonic gap equation when the quark sector is ignored \cite{Feuchter2004, Feuchter2004a}. The quark contributions to the gluon gap equation (\ref{Gl: Gapglomega}) arise from the two-loop contributions of the free Dirac Hamiltonian (\ref{Gl: DiracHamiltonian}) to the vacuum energy
\begin{align}
I_{V V}^{\mathrm{YM}}(k) &= 2 g^2 \int \da^3 p \, V^2(\vp, \vk - \vp) X(\vp, \vk - \vp) p P^2(p) P(|\vk - \vp|) \left(1 - S^2(p) + 2 S(p) S(|\vk - \vp|)\right) \label{Gl: FrDiracGluonV} \\
I_{W W}^{\mathrm{YM}}(k) &= 2 g^2 \int \da^3 p \, W^2(\vp, \vk - \vp) Y(\vp, \vk - \vp) p P^2(p) P(|\vk - \vp|) \left(1 - S^2(p) - 2 S(p) S(|\vk - \vp|)\right) \label{Gl: FrDiracGluonW}
\end{align}
\begin{figure}%
\centering%
\parbox{0.125\linewidth}{%
\centering%
\includegraphics[width=\linewidth]{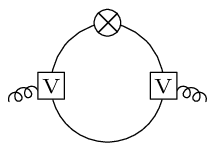} \\%
(a) %
}%
\hspace{0.1\linewidth}%
\parbox{0.125\linewidth}{%
\centering%
\includegraphics[width=\linewidth]{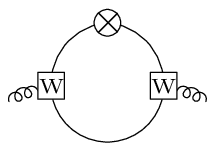} \\%
(b) %
}%
\caption{Diagrammatic representation of the contributions of the free Dirac Hamiltonian to the gluon gap equation, (a) eq.~(\ref{Gl: FrDiracGluonV}) and (b) eq.~(\ref{Gl: FrDiracGluonW}).}%
\label{Abb: FeynmanFrDiracGluon}%
\end{figure}%
illustrated in fig.~\ref{Abb: FeynmanFrDiracGluon}, and from the quark-gluon coupling in the Dirac Hamiltonian (\ref{Gl: DiracHamiltonian}) resulting in the one-loop contributions
\begin{align}
I_{V \mathrm{Q}}^{\mathrm{YM}}(k) &= -2 g^2 \int \da^3 p \, V(\vp, \vk - \vp) X(\vp, \vk - \vp) P(p) P(|\vk - \vp|) \left(1 + S(p) S(|\vk - \vp|)\right) \label{Gl: KopplungstermGluonV} \\
I_{W \mathrm{Q}}^{\mathrm{YM}}(k) &= -2 g^2 \int \da^3 p \, W(\vp, \vk - \vp) Y(\vp, \vk - \vp) P(p) P(|\vk - \vp|) \left(S(p) + S(|\vk - \vp|)\right) \label{Gl: KopplungstermGluonW}
\end{align}
\begin{figure}%
\centering%
\parbox{0.125\linewidth}{%
\centering%
\includegraphics[width=\linewidth]{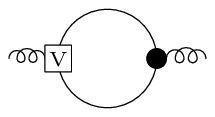} \\%
(a) %
}%
\hspace{0.1\linewidth}%
\parbox{0.125\linewidth}{%
\centering%
\includegraphics[width=\linewidth]{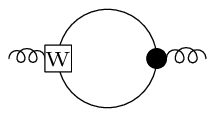} \\%
(b) %
}%
\caption{Diagrammatic representation of the contributions of the quark-gluon coupling in the Dirac Hamiltonian to the gluon gap equation, (a) eq.~(\ref{Gl: KopplungstermGluonV}) and (b) eq.~(\ref{Gl: KopplungstermGluonW}).}%
\label{Abb: FeynmanKopplungstermGluon}%
\end{figure}%
which are diagrammatically illustrated in fig.~\ref{Abb: FeynmanKopplungstermGluon}. Note that the quark contribution to the kinetic energy of the transversal gluons $\langle H_{\mathrm{YM}}^E \rangle$ [the last two terms in eq.~(\ref{Gl: Kinen1})] does not contribute to the gluonic gap equation, while it does contribute to the quark gap equation (\ref{Gl: Gapgleichung}).

Neglecting the quark-gluon coupling in the ansatz for the vacuum wave functional, $V = W = 0$, the gluonic gap equation (\ref{Gl: Gapglomega}) agrees with the result from pure Yang--Mills theory, see eq.~(\ref{Gl: GapglomegaYM}).

A fully unquenched calculation would now necessitate to solve the system of coupled integral equations for the scalar quark kernel $S$  and the bosonic kernel $\omega$, (\ref{Gl: Gapgleichung}) and (\ref{Gl: Gapglomega}), while the vectorial kernels $V$ and $W$ can simply be expressed in terms of these two kernels, see (\ref{Gl: VKern}) and (\ref{Gl: WKern}). This goes beyond the scope of the present paper and will be subject of further research. Here we focus on the quark sector and confine ourselves mainly to a quenched calculation using the previously obtained results for the Yang--Mills sector as input.\footnote{We will, however, consider the unquenching of the gluon propagator, see section \ref{Abschn: RenGluongleichung}.} To be more precise, for the gluon energy $\omega$ we will use Gribov's formula (\ref{Gl: Gribov}) which nicely fits the (quenched) lattice data. Furthermore, the true Coulomb potential (\ref{Gl: Coulombkern3}) obtained in the variational approach \cite{ERS2007} can be approximated to good accuracy by the sum of a linearly rising potential and an ordinary Coulomb potential, which reads in momentum space
\beq
V_{\mathrm{C}}(p) = \frac{8 \pi \sigma_{\mathrm{C}}}{p^4} - \frac{4 \pi \alpha_{\mathrm{S}}}{p^2} = V_{\mathrm{C}}^{\mathrm{IR}}(p) + V_{\mathrm{C}}^{\mathrm{UV}}(p). \label{Gl: Coulombkern2}
\eeq
In accordance with lattice calculations we choose the Coulomb string tension as
\beq
\sigma_{\mathrm{C}} = 2 \sigma \label{Gl: Stringtension}
\eeq
with $\sigma = (440 \, \mathrm{MeV})^2$ being the Wilson string tension.

\section{UV analysis and renormalization} \label{Abschn: UV}

In the quenched approximation only the quark gap equation for the scalar kernel $S$ remains to be solved while the vector kernels are explicitly available in terms of $\omega$ and $S$, see eqs.~(\ref{Gl: VKern}) and (\ref{Gl: WKern}). The loop integrals in the quark gap equation contain UV divergences and the UV analysis carried out in appendix \ref{Anh: UV} reveals the following UV behavior: All loop integrals on the r.h.s. of the quark gap equation (\ref{Gl: Gapgleichung}) contain linear and logarithmic UV divergences except for the one arising from the Coulomb term, $I_{\mathrm{C}}^{\mathrm{Q}}$ [eq.~(\ref{Gl: CoulombQQGapgl})], which is only logarithmically divergent. The logarithmic divergence arises here from the Coulomb potential $V_{\mathrm{C}}^{\mathrm{UV}}$, see eq.~(\ref{Gl: Coulombkern2}). The linear divergences are known to cancel by gauge invariance, at least in covariant perturbation theory. In fact, adding the linear divergent contributions in the quark gap equation (\ref{Gl: Gapgleichung}) we find that they exactly cancel. To be more precise, this strict cancellation is a direct consequence of the inclusion of the vector kernel $W$ into the ansatz for the vacuum wave functional [eqs.~(\ref{Gl: VakuumfunktionalQ}) and (\ref{Gl: VakuumfunktionalQ1})]. Without this kernel, the quark gap equation would contain both a logarithmic and a linear UV divergence, which makes the inclusion of $W$ not only a quantitative but also a qualitative improvement of the variational ansatz. Due to the strict cancellation of the linear UV divergences, the quark gap equation (\ref{Gl: Gapgleichung}) contains only logarithmic divergences. The logarithmically divergent terms of the gap equation (\ref{Gl: Gapgleichung}) sum up to
\beq
\frac{C_{\mathrm{F}}}{3 \pi^2} g^2 k S(k) \ln \frac{\Lambda}{\mu} \label{Gl: UVDivGapgl}
\eeq
where $\Lambda$ is the UV cutoff and $\mu$ is an, so far, arbitrary momentum scale. As can be seen from this expression, the remaining logarithmic UV divergences are multiplied by a factor $g^2$ and can hence be absorbed in a renormalized coupling
\beq
\widetilde{g}^2(\mu) = g^2 \ln \frac{\Lambda}{\mu} \,. \label{Gl: Renormierung}
\eeq
The renormalized coupling $\widetilde{g}$ is assumed to be finite so that the bare coupling $g$ vanishes for $\Lambda \to \infty$ like $1/\sqrt{\ln \Lambda}$. Ignoring terms which vanish for $\Lambda \to \infty$ with $\widetilde{g}^2(\mu) = \mathrm{const.} < \infty$ the renormalized quark gap equation reduces to
\beq
k S(k) \left[1 - \frac{C_{\mathrm{F}}}{3 \pi^2} \widetilde{g}^2(\mu)\right] = \frac{C_{\mathrm{F}}}{2} \int \da^3 p \, V_{\mathrm{C}}^{\mathrm{IR}}(|\vp - \vk|) P(p) \left[S(p) \bigl(1 - S^2(k)\bigr) - S(k) \bigl(1 - S^2(p)\bigr) \hp \cdot \hk\right] \,. \label{Gl: Gapgleichung2}
\eeq
This equation is considerably simpler than the unrenormalized gap equation (\ref{Gl: Gapgleichung}). The whole effect of the quark gluon coupling as well as the UV part of the Coulomb potential $V_{\mathrm{C}}^{\mathrm{UV}}$ [eq.~(\ref{Gl: Coulombkern2})] are now captured by the term multiplied by the renormalized coupling constant. In fact, the quark-gluon coupling and $V_{\mathrm{C}}^{\mathrm{UV}}$ equally contribute to the renormalized gap equation. When the renormalized coupling constant $\widetilde{g}$ is set to zero eq.~(\ref{Gl: Gapgleichung2}) reduces to the gap equation of ref.~\cite{Adler1984}.

If the confining part of the non-Abelian Coulomb potential is neglected, $V_{\mathrm{C}}^{\mathrm{IR}} = 0$, the gap equation (\ref{Gl: Gapgleichung2}) has only the trivial solution $S = 0$. This result is in agreement with the empirical findings of ref.~\cite{Pak2013} that there is no spontaneous breaking of chiral symmetry when $V_{\mathrm{C}}^{\mathrm{IR}}$ is neglected. Although the quark-gluon coupling term alone cannot \textit{trigger} spontaneous breaking of chiral symmetry it considerably increases the strength of the symmetry breaking as can be read off from eq.~(\ref{Gl: Gapgleichung2}) and as will later on be confirmed by the numerical calculation. The r.h.s. of eq.~(\ref{Gl: Gapgleichung2}) is independent of $\widetilde{g}$, while on the l.h.s. a non-zero $\widetilde{g}$ reduces the factor multiplying $k S(k)$ and thus increases $S$, which also increases the quark condensate (see below).

Renormalizing the static quark propagator (\ref{Gl: StatProp1}) in the same way as the gap equation (\ref{Gl: Gapgleichung}), see appendix \ref{Anh: UV}, we find for the renormalized propagator
\beq
G_{\mathrm{ren}}(\vp) = \frac{1}{2} \left[1 - \frac{C_{\mathrm{F}}}{8 \pi^2} \widetilde{g}^2(\mu)\right] P(p) \Bigl[\Bigl(1 - S^2(p)\Bigr) \valpha \cdot \hp + 2 S(p) \beta\Bigr] = \left[1 - \frac{C_{\mathrm{F}}}{8 \pi^2} \widetilde{g}^2(\mu)\right] \frac{\valpha \cdot \vp + \beta M(p)}{2 \sqrt{p^2 + M^2(p)}} \,. \label{Gl: StatPropRen}
\eeq
Like the unrenormalized propagator (\ref{Gl: StatProp1}) it has the quasi-particle form (\ref{Gl: StatProp3}) with an effective (running) mass $M$ given by eq.~(\ref{Gl: ADMasse}) and the constant field renormalization factor
\beq
Z(p) = 1 - \frac{C_{\mathrm{F}}}{8 \pi^2} \widetilde{g}^2(\mu) \,.
\eeq
From the renormalized quark propagator (\ref{Gl: StatPropRen}) we find the renormalized quark condensate
\beq
\langle \bar{\psi}(\vx) \psi(\vx) \rangle = -4 \Ncc \left[1 - \frac{C_{\mathrm{F}}}{8 \pi^2} \widetilde{g}^2(\mu)\right] \int \da^3 p \, P(p) S(p) \,. \label{Gl: ChiralesKondensat2}
\eeq
Obviously, the quark condensate vanishes for the trivial solution $S = 0$ of the gap equation. Furthermore, from the renormalized quark propagator (\ref{Gl: StatPropRen}) we can conclude that the \mbox{(anti-)particle} occupation numbers are given by\footnote{Note that there is no summation over spin and color indices on the l.h.s. and that a spatial volume factor $\deltabar^3(0)$ has been ignored.}
\begin{align}
\langle {a_{\mathrm{ren}}^{s, m}}^{\dagger}(\vp) a_{\mathrm{ren}}^{s, m}(\vp) \rangle &= P(p) S^2(p) \label{Gl: BesetzungszahlQ}
\intertext{and}
\left(1 - \langle {b_{\mathrm{ren}}^{s, m}}^{\dagger}(\vp) b_{\mathrm{ren}}^{s, m}(\vp) \rangle\right) &=  P(p) \,, \label{Gl: BesetzungszahlAQ}
\end{align}
respectively, where $a$ ($b$) is the annihilation operator for a (anti-)quark in momentum space, see appendix \ref{Anh: Impulsdarstellung}. Note that we have, at this point, replaced the quark field $\psi$ in terms of the field renormalization factor $Z$ by the renormalized quantity $\psi_{\mathrm{ren}}$ in order to obtain the physical occupation numbers, i.e. $a_{\mathrm{ren}}^{\dagger}$, $b_{\mathrm{ren}}^{\dagger}$ generate properly normalized one-particle states. Furthermore, from eqs.~(\ref{Gl: PFaktor}), (\ref{Gl: BesetzungszahlQ}) and (\ref{Gl: BesetzungszahlAQ}) one can read off immediately that $\langle a_{\mathrm{ren}}^{\dagger} a_{\mathrm{ren}} \rangle = \langle b_{\mathrm{ren}}^{\dagger} b_{\mathrm{ren}} \rangle$ holds.

The momentum scale $\mu$ introduced above in the context of the renormalized coupling (\ref{Gl: Renormierung}) is completely arbitrary and can be chosen at will. However, in the quark gap equation (\ref{Gl: Gapgleichung2}) there is a physical scale inherited from the gluon sector: the Coulomb string tension $\sigma_{\mathrm{C}}$, see eq.~(\ref{Gl: Coulombkern2}). It is convenient to identify the arbitrary scale $\mu$ with $\sqrt{\sigma_{\mathrm{C}}}$. This choice of $\mu$ removes the Coulomb string tension from the gap equation when the latter is expressed in dimensionless variables. The only undetermined quantity in our variational equations of motion is then the renormalized coupling $\widetilde{g}$ at the scale $\mu = \sqrt{\sigma_{\mathrm{C}}}$.

\section{Physical implications of the quark-gluon coupling} \label{Abschn: Skalierung}

Below we investigate the immediate consequences of the quark-gluon coupling in our approach on both the quark and the gluon sector.

\subsection{Scaling properties of the quark sector}

For the numerical solution of the gap equation (\ref{Gl: Gapgleichung2}) it is more convenient to express it in terms of the mass function $M$ [eq.~(\ref{Gl: ADMasse})]. This yields
\beq
M(k) \left[1 - \frac{C_{\mathrm{F}}}{3 \pi^2} \widetilde{g}^2(\mu)\right] = 4 \pi C_{\mathrm{F}} \int \da^3 p \, \frac{M(p) - M(k) \frac{\vp \cdot \vk}{{k}^2}}{|\vp - \vk|^4 \sqrt{p^2 + M^2(p)}} \,. \label{Gl: GapgleichungM}
\eeq
Note that the two-loop approximation breaks down when the renormalized coupling constant exceeds the critical value
\beq
\widetilde{g}_{\mathrm{c}} = \sqrt{\frac{3 \pi^2}{C_{\mathrm{F}}}} = \sqrt{\frac{6 \pi^2 \Ncc}{\Ncc^2 - 1}} \, . \label{Gl: KritischeKopplung}
\eeq
However, the critical values $\widetilde{g}_{\mathrm{c}} \approx 6.28$ for $SU(2)$ and $\widetilde{g}_{\mathrm{c}} \approx 4.71$ for $SU(3)$ are well above the value required to reproduce the phenomenological value of the quark condensate, see eq.~(\ref{Gl: EffKopplung}) below, so that the collaps does not occur for realistic values of the renormalized coupling.

Denoting the IR limit of the mass function by $m = M(p = 0)$ and the respective quantities for vanishing coupling by a superscript ``0'', we find from the gap equation (\ref{Gl: GapgleichungM}) the following, very useful scaling properties:
\beq
M(p) = \frac{m}{m^0} M^0(m^0 p / m) \label{Gl: Massenfkt3} \, ,
\eeq
where
\beq
m = \frac{m^{0}}{\sqrt{1 - \frac{C_{\mathrm{F}} \widetilde{g}^2(\mu)}{3 \pi^2}}}
\eeq
holds. From these relations follows that it is sufficient to solve the gap equation (\ref{Gl: GapgleichungM}) only for the case of a vanishing coupling $\widetilde{g} = 0$ yielding the function $M^0$. The solution $M$ of eq.~(\ref{Gl: GapgleichungM}) for arbitrary coupling $\widetilde{g}$ is then obtained from the scaling relation (\ref{Gl: Massenfkt3}). By means of this relation also the quark condensate $\langle \bar{\psi} \psi \rangle$ with quark-gluon coupling included can be related to that for $\widetilde{g} = 0$, $\langle \bar{\psi} \psi \rangle^0$: Using eq.~(\ref{Gl: ADMasse}) to trade the scalar kernel $S$ in eq.~(\ref{Gl: ChiralesKondensat2}) for the mass function $M$ and using subsequently the scaling relation (\ref{Gl: Massenfkt3}) we find the following scaling relation for the quark condensate
\beq
\langle \bar{\psi}(\vx) \psi(\vx) \rangle = \frac{1 - \frac{C_{\mathrm{F}}}{8 \pi^2} \widetilde{g}^2(\mu)}{\sqrt{1 - \frac{C_{\mathrm{F}}}{3 \pi^2} \widetilde{g}^2(\mu)}^3} \langle \bar{\psi}(\vx) \psi(\vx) \rangle^0 \,.
\eeq
Inserting here for the renormalized quark-gluon coupling $\widetilde{g}$ at the scale $\mu = \sqrt{\sigma_{\mathrm{C}}}$ the value (\ref{Gl: EffKopplung}) determined in section \ref{Abschn: Numerik}, we find the relation
\beq
\langle \bar{\psi}(\vx) \psi(\vx) \rangle = (1.42)^3 \langle \bar{\psi}(\vx) \psi(\vx) \rangle^0 \label{Gl: ChiralesKondensat5}
\eeq
thus the quark-gluon coupling increases the quark condensate by $42 \%$.

\subsection{Unquenching the gluon propagator} \label{Abschn: RenGluongleichung}

\begin{figure}
\centering
\includegraphics[width=0.4\linewidth]{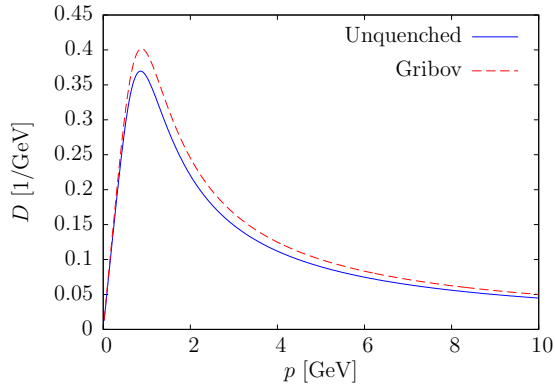}
\caption{The static gluon propagator $D(p) = 1/ 2 \omega(p)$ as given by (\ref{Gl: Gapglomega3}) for an effective coupling of $\widetilde{g}(\sqrt{\sigma_{\mathrm{C}}}) = 3.586$ (full line) and for the Gribov formula (dashed line).}
\label{Abb: Gluonenergie}
\end{figure}

Below we solve the unquenched gluon gap equation (\ref{Gl: Gapglomega}) in a simplified way. This equation has the generic form
\beq
\omega^2(k) = \omega_{\mathrm{YM}}^2(k) + \omega_{\mathrm{Q}}^2(k) \, , \label{Gl: Gapglomega1}
\eeq
where $\omega_{\mathrm{YM}}$, defined by eq.~(\ref{Gl: GapglomegaYM}), is the expression one finds from the pure Yang--Mills sector \cite{Feuchter2004, Feuchter2004a} and
\beq
\omega_{\mathrm{Q}}^2(k) = I_{V V}^{\mathrm{YM}}(k) + I_{W W}^{\mathrm{YM}}(k) + I_{V \mathrm{Q}}^{\mathrm{YM}}(k) + I_{W \mathrm{Q}}^{\mathrm{YM}}(k) \label{Gl: GapglomegaQ}
\eeq
is the quark contribution, which consists of several one-loop terms defined by eqs.~(\ref{Gl: FrDiracGluonV}) to (\ref{Gl: KopplungstermGluonW}). Assuming for the scalar quark kernel $S$ a sufficiently fast vanishing UV asymptotics one can work out the UV behavior of these loop integrals (see appendix \ref{Anh: UV}) and finds that they are quadratic plus logarithmic divergent
\beq
\label{1180-2}
\omega_{\mathrm{Q}}^2(k) = -\frac{g^2}{6 \pi^2} \left[\Lambda^2 - \frac{1}{2} \bigl(\omega^2(k) + k^2\bigr) \ln\frac{\Lambda}{\mu}\right] + \mbox{finite terms} \, .
\eeq
The quadratic UV divergence is thus independent of the external momentum $k$. Such an UV divergence also occurs in the Yang--Mills sector, i.e. in $\omega_{\mathrm{YM}}$ (\ref{Gl: GapglomegaYM}) and can therefore be removed by the known renormalization procedure of the Yang--Mills sector, i.e. by the counterterm $\sim A^2$ \cite{ERSS2008}. Like in the quark gap equation, the (logarithmic) UV divergence is here multiplied by a factor $g^2$ and can be absorbed into the renormalized coupling constant $\widetilde{g}$ (\ref{Gl: Renormierung}). Droping all terms which vanish for $\Lambda \to \infty$ for finite $\widetilde{g}$, we arrive at the unquenched renormalized gluon gap equation
\beq
\omega^2(k) = \bar{\omega}_{\mathrm{YM}}^2(k) + \frac{\widetilde{g}^2(\mu)}{12 \pi^2} \bigl(\omega^2(k) + k^2\bigr) \, , \label{Gl: Gapglomega2}
\eeq
where $\bar{\omega}_{\mathrm{YM}}$ is the renormalized version of $\omega_{\mathrm{YM}}$ (\ref{Gl: GapglomegaYM}), see ref.~\cite{ERSS2008}. The equation (\ref{Gl: Gapglomega2}) can be solved for $\omega$ yielding
\beq
\omega^2(k) = \left[\frac{\widetilde{g}^2(\mu)}{12 \pi^2} k^2 + \bar{\omega}_{\mathrm{YM}}^2(k) \right] \left[1 - \frac{\widetilde{g}^2(\mu)}{12 \pi^2}\right]^{- 1}. \label{Gl: Gapglomega3}
\eeq
To obtain a first estimate of the effect of unquenching on the gluon propagator we use for the quenched (renormalized) gluon energy $\bar{\omega}_{\mathrm{YM}}$ the Gribov formula (\ref{Gl: Gribov}). The resulting gluon propagator $1/ (2 \omega)$ is shown in fig.~\ref{Abb: Gluonenergie} together with the quenched Gribov propagator. The unquenching leads to a decrease of the gluon propagator in the mid-momentum regime but the effect of unquenching  is small. A similar result was obtained in ref.~\cite{Pak2013} where the unquenching was done in the same way.

\section{Numerical results} \label{Abschn: Numerik}

By the above derived scaling relation (\ref{Gl: Massenfkt3}) we have reduced the numerical calculation to the case with vanishing quark-gluon coupling $\widetilde{g} = 0$. The numerical solution of the remaining quark gap equation (\ref{Gl: GapgleichungM}) (with $\widetilde{g} = 0$) can be performed by standard methods, see e.g. ref.~\cite{Watson2012}. The resulting mass function for vanishing coupling can be nicely fitted by
\beq
M_{\mathrm{fit}}^{0}(p) = \frac{m^{0}}{1 + a p^{2 A} + 2 b p^{2 B}} \label{Gl: Fitfunktion}
\eeq
with the optimized fit parameters
\begin{alignat}{3}
m^0 &= 0.165508 \, &\quad \quad a &= 4.59935 \, & \quad \quad b &= 1.26155 \nonumber \\
A &= 0.985198 \, &\quad \quad B &= 2.27051 \,. & &
\end{alignat}
From this fit we read-off the UV power-law behavior $M(p \to \infty) \sim p^{-4.54}$ which, according to (\ref{Gl: Massenfkt3}), holds for arbitrary strengths of the quark-gluon coupling. This UV exponent differs somewhat from the value $(-4)$ found in refs.~\cite{Watson2012, Pak2013}\footnote{The numerical extraction of the UV exponent from the solution is quite subtle and depends sensitively on the actual momentum interval used.} which, however, has almost no effect on the quantities calculated below since the UV part is highly suppressed.

\begin{figure}%
\centering%
\parbox{0.4\linewidth}{%
\centering%
\includegraphics[width=\linewidth]{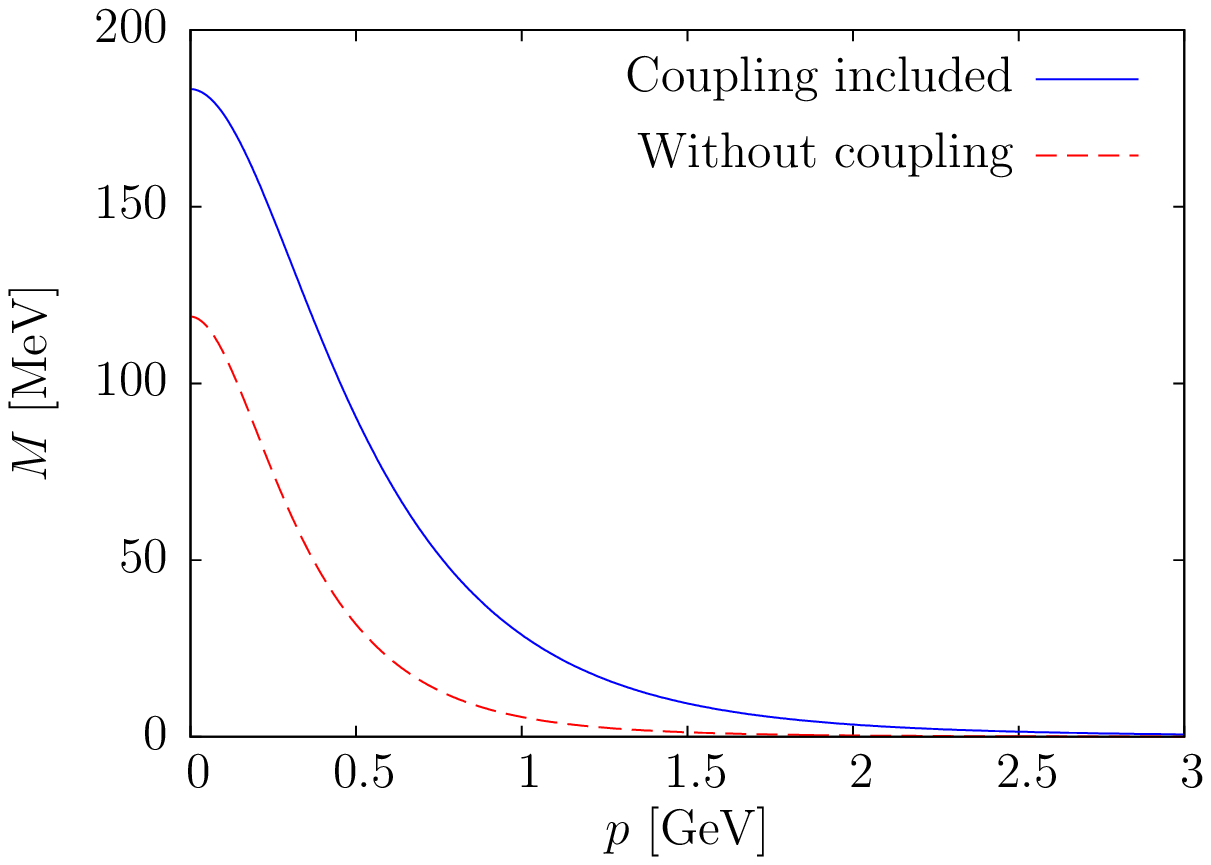} \\%
(a) %
}%
\hspace{0.1\linewidth}%
\parbox{0.4\linewidth}{%
\centering%
\includegraphics[width=\linewidth]{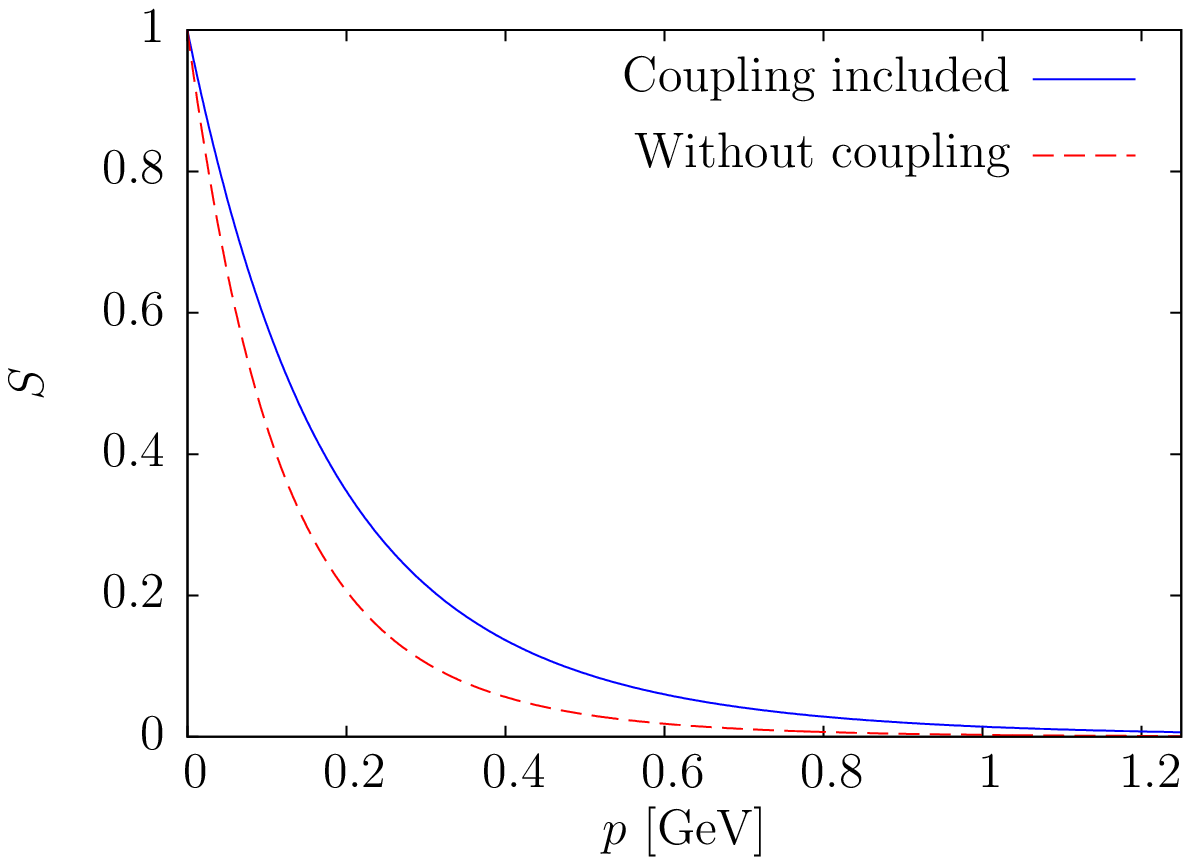} \\%
(b) %
}%
\caption{(a) Mass function for an effective coupling strength of $\widetilde{g}(\sqrt{\sigma_{\mathrm{C}}}) = 3.586$ (full curve) compared to the result obtained with a BCS-ansatz neglecting the coupling (dashed curve). The Coulomb string tension is set to $\sigma_{\mathrm{C}} = 2 \sigma$. (b) Scalar kernel for an effective coupling strength of $\widetilde{g}(\sqrt{\sigma_{\mathrm{C}}}) = 3.586$ (full curve) compared to the result neglecting the coupling (dashed curve).}%
\label{Abb: Massenfunktion}%
\end{figure}

We adjust the renormalized effective coupling constant $\widetilde{g}$ such that the phenomenological value
\beq
\langle \bar{\psi}(\vx) \psi(\vx) \rangle_{\mathrm{phen}} = (-235 \, \mathrm{MeV})^3 \label{Gl: ChiralesKondensatphaen}
\eeq
is reproduced. This requires for $\sigma_{\mathrm{C}} = 2 \sigma$ an effective coupling constant of
\beq
\widetilde{g}(\sqrt{\sigma_{\mathrm{C}}}) = 3.586 \, . \label{Gl: EffKopplung}
\eeq
When the coupling of the quarks to the gluons is neglected one obtains (for the same value of the Coulomb string tension) for the quark condensate the substantially smaller value
\beq
\langle \bar{\psi}(\vx) \psi(\vx) \rangle^0 = (-165 \, \mathrm{MeV})^3  \, .\label{Gl: ChiralesKondensatAD}
\eeq
Using for the renormalized coupling constant $\widetilde{g}$ the value (\ref{Gl: EffKopplung}) obtained from the phenomenological quark condensate one finds the solution of the gap equation (\ref{Gl: GapgleichungM}) shown in fig.~\ref{Abb: Massenfunktion} (a). The IR mass is given by $m \approx 210 \, \mathrm{MeV}$. For sake of comparison, fig.~\ref{Abb: Massenfunktion} (a) shows also the solution $M^0$ of the gap equation for vanishing quark-gluon coupling $\widetilde{g} = 0$. The effective mass is then considerably smaller. This refers in particular to its IR value $m^0 \approx 120 \, \mathrm{MeV}$. Fig.~\ref{Abb: Massenfunktion} (b) shows the corresponding results for the scalar kernel $S$, which can be calculated from eq.~(\ref{Gl: ADMasse}) once $M$ is known. The IR value $S(p = 0) = 1$ is obtained independently of the value of the quark-gluon coupling constant $\widetilde{g}$ as one can also extract analytically from the gap equation. Neglecting the quark-gluon coupling considerably decreases $S$ in the mid-momentum regime. It is this momentum regime which dominantly contributes to the spontaneous breaking of chiral symmetry, i.e. to the quark condensate [see eq.~(\ref{Gl: ChiralesKondensat2})]. This is also found in lattice invesitgations \cite{Suganuma2010}.

\begin{figure}%
\centering%
\includegraphics[width=0.4\linewidth]{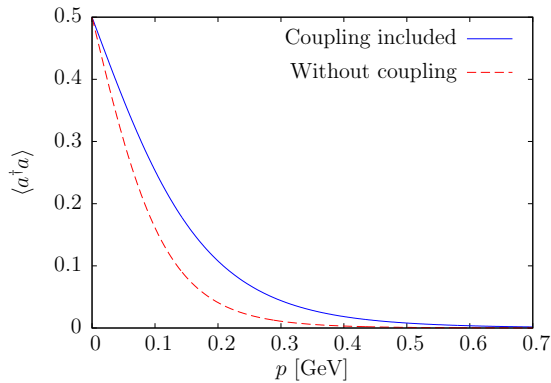}
\caption{Occupation number of quarks according to eq.~(\ref{Gl: BesetzungszahlQ}). The straight line shows the results for the quark-gluon coupling $\widetilde{g}(\sqrt{\sigma_{\mathrm{C}}}) = 3.586$ while the dashed one is for $\widetilde{g} = 0$.}%
\label{Abb: Besetzungszahl}%
\end{figure}

With the scalar kernel $S$ at our disposal we are also able to determine the quark occupation number eq.~(\ref{Gl: BesetzungszahlQ}). The resulting curve is shown in fig.~\ref{Abb: Besetzungszahl}. One can clearly observe that the inclusion of the quark-gluon coupling significantly increases the occupation number in the mid-momentum regime. Note that the same result holds for the occupation number of anti-quarks [see eq.~(\ref{Gl: BesetzungszahlAQ}) and the discussion following thereafter].

\begin{figure}
\centering%
\parbox{0.47\linewidth}{%
\centering%
\includegraphics[trim = 60mm 0mm 60mm 0mm, width=\linewidth]{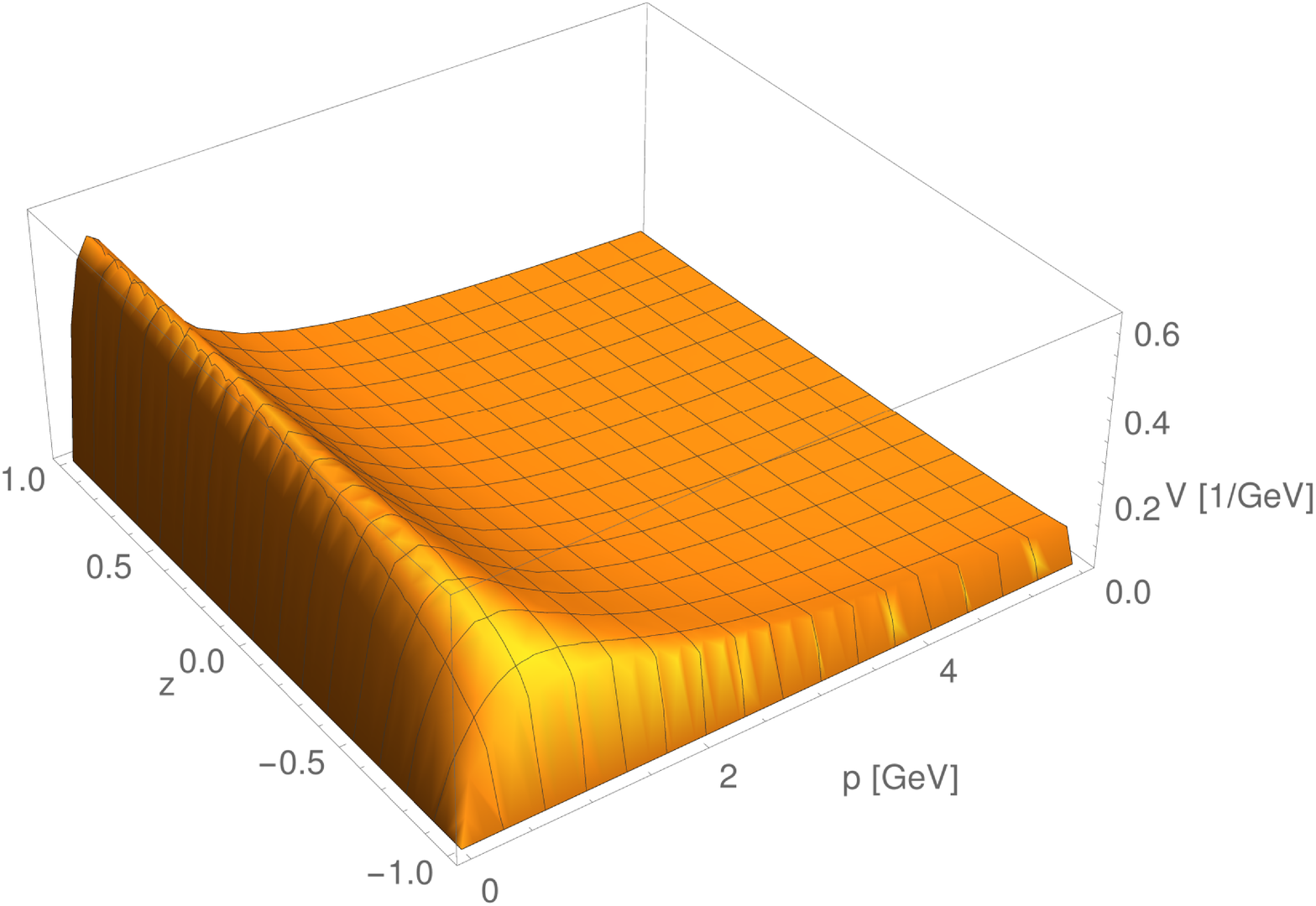} \\%
(a) %
}%
\hfill%
\parbox{0.47\linewidth}{%
\centering%
\includegraphics[trim = 60mm 0mm 60mm 0mm, width=\linewidth]{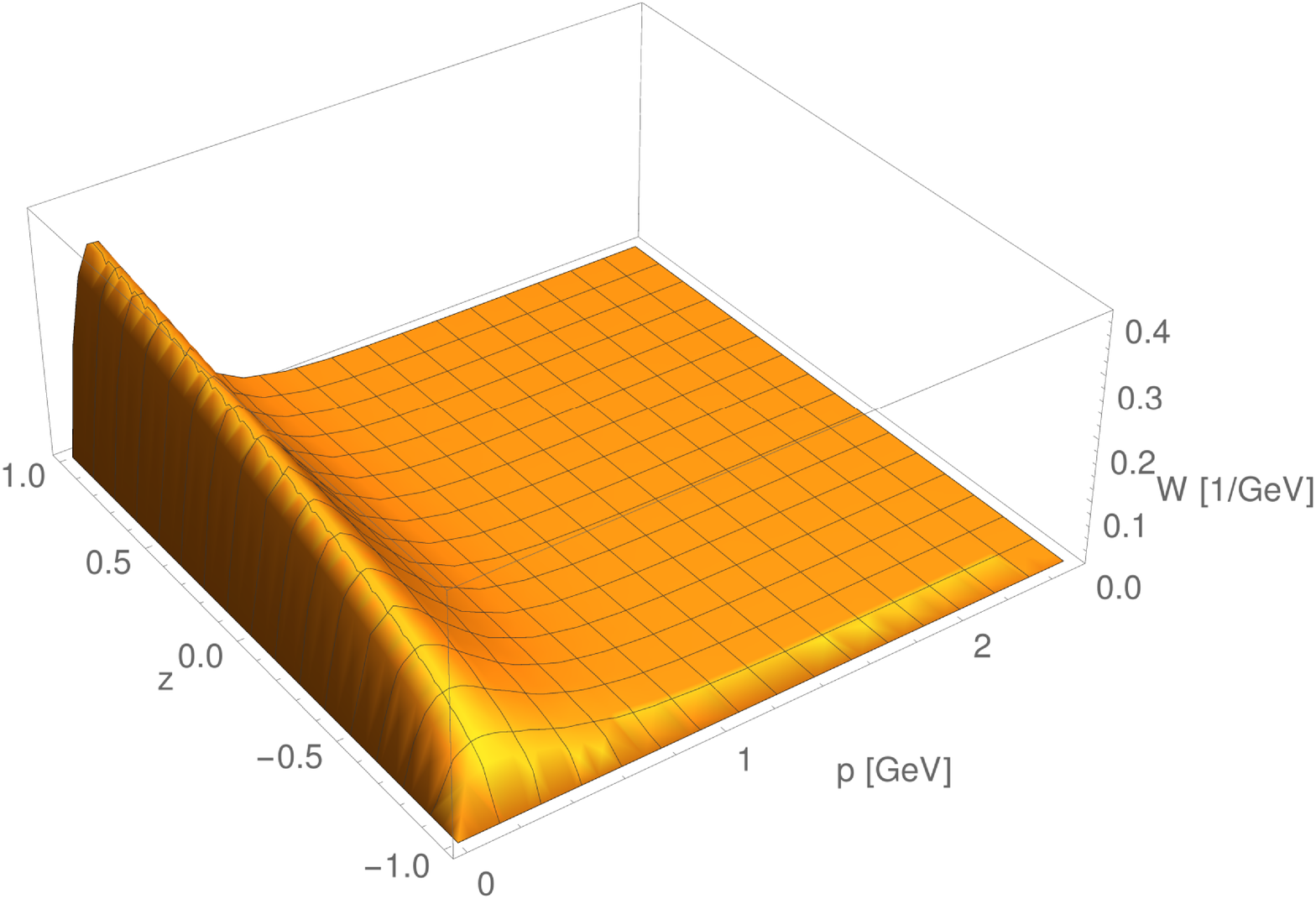} \\%
(b) %
}%
\caption{The vector kernels (a)  $V(\vp, \vq)$ and (b) $W (\vp, \vq)$  for an effective coupling strength of $\widetilde{g}(\sqrt{\sigma_{\mathrm{C}}}) = 3.586$ as function of the modulus $p = q$ and $z = \cos \sphericalangle(\vp,\vq)$.}
\label{Abb: VKern}
\end{figure}

Finally, we explicitly evaluate the vector kernels  $V$ [eq.~(\ref{Gl: VKern})]  and $W$ [eq.~(\ref{Gl: WKern})] using the scalar kernel $S$ obtained above and the Gribov formula (\ref{Gl: Gribov}) as input for the gluon energy. The results are plotted in fig.~\ref{Abb: VKern} for the case that the moduli of the two momentum arguments agree. Although the shape of both form factors is similar they differ in the size (note the different scales used in figs.~\ref{Abb: VKern} (a) and (b), respectively): $V$ is considerably larger than $W$. Furthermore, for $p = q$ the kernel $W(\vp, \vq)$ drops off more rapidly in the UV ($p = q \to \infty$) than $V(\vp, \vq)$. This can be also read off from the explicit analytic expressions given in eqs.~(\ref{Gl: VKern}) and (\ref{Gl: WKern}). From these expressions it also follows that $W$ is, in general, much more sensitive to the detailed behavior of the scalar kernel $S$ than $V$ (which is not surprising for $W$ being purely non-perturbative). Keeping one momentum argument, say $q$, fixed both vector form factors vanish like $1/p$ for $p \to \infty$.

\section{Summary and conclusions} \label{Abschn: Summary}

In this paper the variational approach to QCD originally developed in refs.~\cite{Feuchter2004, Feuchter2004a, ERS2007} for the Yang--Mills sector and extended in refs.~\cite{Pak2012a, Pak2013} to full QCD has been further developed and essentially improved in several regards: First we included the UV part $V_{\mathrm{C}}^{\mathrm{UV}}$ of the color Coulomb potential in the treatment of the quark sector which was not the case in refs.~\cite{Pak2012a, Pak2013}. This additional contribution turns out to have the same effect on the renormalized gap equation as the quark-gluon coupling. Second, we have abandoned a simplifying approximation used in ref.~\cite{Pak2013} and calculated the vacuum energy density consistently to two-loop order. The resulting expressions are consistent with the results obtained by using Dyson--Schwinger equations to carry out the variational approach \cite{CR201Xa} and also with perturbation theory \cite{CR2015a}. Third, we have included an additional Dirac structure (with a new variational kernel) into the fermionic vacuum wave functional, thereby enlarging the space of trial states which improves the variational results. This new Dirac structure in the quark wave functional can be motivated by treating the quark-gluon coupling of the QCD Hamiltonian in perturbation theory on top of the non-trivial BCS-state obtained in a mean-field treatment of the non-Abelian Coulomb interaction. By including this new quark-gluon coupling term in the trial vacuum state all linear UV divergences are eliminated from the quark gap equation. The remaining logarithmic UV divergence has been absorbed into a renormalized coupling constant. In the resulting renormalized gap equation the total quark-gluon coupling becomes manifest in a rescaling of the effective (momentum dependent) quark mass. By using this scaling property physical quantities like the quark condensate can be related to their values when the quark-gluon coupling is switched off. In this way we could analytically show that the inclusion of the quark-gluon coupling in the vacuum wave functional leads to a substantial increase in the quark condensate and the effective quark mass. In our Hamiltonian approach to QCD in Coulomb gauge the quark sector inherits a scale from the gluon sector. This is given by the Coulomb string tension $\sigma_{\mathrm{C}}$, which also determines the Gribov mass $M_{\mathrm{G}} \approx \sigma_{\mathrm{C}}$. Choosing the renormalized coupling at this scale to reproduce the phenomenological value of the quark condensate, $\langle \bar{\psi} \psi \rangle = (-235 \, \mathrm{MeV})^3$, requires a value of $\widetilde{g}(\sqrt{\sigma_{\mathrm{C}}}) = 3.586$.

Using for the quenched gluon propagator the lattice results, which can be nicely fitted by the Gribov formula, and renormalizing the full unquenched gluon gap equation consistently with the quark gap equation, we have shown that the unquenched gluon propagator is somewhat reduced in the mid-momentum regime compared to the quenched one but the effect is quite small.

The results obtained in the present paper are quite encouraging for further applications of the present approach. We plan to extend it to finite temperatures and baryon densities to study the deconfinement and chiral phase transitions. In a first application we will then calculate the quark contribution to the effective potential of the Polyakov loop extending the  approach of ref.~\cite{RH2013} to full QCD.

\section*{Acknowledgments}

The authors thank E. Ebadati for numerical support and discussions, and M. Quandt for a critical reading of the manuscript and useful comments. This work was supported by Deutsche Forschungsgemeinschaft (DFG) under Contract No. DFG-Re856/10-1.

\appendix

\section{Momentum representation} \label{Anh: Impulsdarstellung}

Due to translational invariance of the vacuum it is convenient to carry out the actual calculations in momentum space. Thereby, it will be convenient to expand the quark field $\psi$ in terms of the eigenspinors $u$, $v$ of the free Dirac Hamiltonian of chiral quarks
\beq
h_0(\vp) = \vec{\alpha} \cdot \vp \,.
\eeq
We choose the eigenspinors such that they satisfy the eigenvalue equation
\beq
\valpha \cdot \vp \, u^s(\vp) = p u^s(\vp), \quad \quad \valpha \cdot \vp \, v^s(-\vp) = -p v^s(-\vp)
\eeq
and are normalized according to
\begin{subequations}
\begin{align}
{u^s}^{\dagger}(\vp) u^t(\vp) &= {v^s}^{\dagger}(-\vp) v^t(-\vp) = 2 p \delta^{s t} \\
{u^s}^{\dagger}(\vp) v^t(-\vp) &= {v^s}^{\dagger}(-\vp) u^t(\vp) = 0 \,.
\end{align}
\label{Gl: SpinorNormierung}%
\end{subequations}
Here $s = \pm 1$ denotes the double of the spin projection. The spinors satisfy the relations
\begin{subequations}
\begin{align}
\beta u^s(\vp) &= s v^s(-\vp) \\
\beta v^s(-\vp) &= s u^s(\vp)
\end{align}
\end{subequations}
\begin{subequations}
\begin{align}
\gamma_5 u^s(\vp) &= s u^s(\vp) \\
\gamma_5 v^s(-\vp) &= -s v^s(-\vp)
\end{align}
\end{subequations}
from which we can conclude
\begin{subequations}
\begin{align}
\bar{u}^s(\vp) u^t(\vp) &\equiv {u^s}^{\dagger}(\vp) \beta u^t(\vp) = 0 \\
\bar{v}^s(-\vp) v^t(-\vp) &= 0.
\end{align}
\end{subequations}
For the outer product of the spinors one finds\footnote{Note that there is no summation over the spin index.}
\begin{subequations}
\begin{align}
u^s(\vp) {u^s}^{\dagger}(\vp) &= \frac{1}{2} p \bigl(\left[\Id + \valpha \cdot \hp\right] + s \gamma_5 \left[\Id + \valpha \cdot \hp\right]\bigr) \\
v^s(-\vp) {v^s}^{\dagger}(-\vp) &= \frac{1}{2} p \bigl(\left[\Id - \valpha \cdot \hp\right] - s \gamma_5 \left[\Id - \valpha \cdot \hp\right]\bigr) \\
u^s(\vp) {v^s}^{\dagger}(-\vp) &= \frac{1}{2} p \bigl(s \beta \left[\Id - \valpha \cdot \hp\right] - \beta \gamma_5 \left[\Id - \valpha \cdot \hp\right]\bigr) \\
v^s(-\vp) {u^s}^{\dagger}(\vp) &= \frac{1}{2} p \bigl(s \beta \left[\Id + \valpha \cdot \hp\right] + \beta \gamma_5 \left[\Id + \valpha \cdot \hp\right]\bigr)
\end{align}
\end{subequations}
while the matrix elements of the Dirac operator are given by
\begin{subequations}
\begin{align}
{u^s}^{\dagger}(\vp) \valpha \cdot \vq u^t(\vp) &= 2 p q \bigl(\hp \cdot \hq\bigr) \delta^{s t} \\
{v^s}^{\dagger}(-\vp) \valpha \cdot \vq v^t(-\vp) &= -2 p q \bigl(\hp \cdot \hq\bigr) \delta^{s t} \\
{u^s}^{\dagger}(\vp) \valpha \cdot \vq v^s(-\vp) &= 0 \\
{v^s}^{\dagger}(-\vp) \valpha \cdot \vq u^s(\vp) &= 0 \,.
\end{align}
\end{subequations}
In terms of these eigenspinors the quark field can be expanded as
\beq
\psi^m(\vx) = \int \da^3 p \, \frac{1}{\sqrt{2 p}} \exp(\ii \vp \cdot \vx) \Bigl(a^{s, m}(\vp) u^s(\vp) + {b^{s, m}}^{\dagger}(-\vp) v^s(-\vp)\Bigr) \label{Gl: EntwicklungFeldop}
\eeq
where $a^s$ ($b^s$) is the annihilation operator of a (anti-)quark in a state with spin projection $s/2 = \pm 1/2$. With the normalization (\ref{Gl: SpinorNormierung}) of the eigenspinors the fermionic anti-commutation relations (\ref{Gl: Antikommutator}) are fulfilled when the $a, b$ satisfy the relations
\beq
\left\{a^{s, m}(\vp), {a^{t, n}}^{\dagger}(\vq)\right\} = \left\{b^{s, m}(\vp), {b^{t, n}}^{\dagger}(\vq)\right\} = \delta^{s t} \delta^{m n} \deltabar^3(\vp - \vq)
\eeq
with all other anti-commutators vanishing. With the expansion (\ref{Gl: EntwicklungFeldop}) and the orthonormality relation of the Dirac eigenmodes our quark vacuum wave functional (\ref{Gl: VakuumfunktionalQ}) aquires the form
\beq
\rvert \phi_{\mathrm{Q}}[A] \rangle = \exp\left(-\int \da^3 p \int \da^3 q \, K^{s t, m n}(\vp, \vq) {a^{s, m}}^{\dagger}(\vp) {b^{t, n}}^{\dagger}(\vq)\right) \rvert 0 \rangle \label{Gl: FermiAnsatz}
\eeq
where the kernel $K$ in momentum space is related to that in the coordinate representation by
\begin{align}
K^{s t, m n}(\vp, \vq) &= \frac{1}{2 \sqrt{p q}} {u^s}^{\dagger}(\vp) \int \dd^3 x \int \dd^3 y \, \exp\bigl(-\ii \vp \cdot \vx - \ii \vq \cdot \vy\bigr) K^{m n}(\vx, \vy) v^t(\vq) \nonumber \\
&= \delta^{m n} \delta^{s t} \deltabar(\vp + \vq) s S(p) + g \frac{1}{2 \sqrt{p q}} t_{m n}^a {u^s}^{\dagger}(\vp) \bigl(V(\vp, \vq) + \beta W(\vp, \vq)\bigr) \alpha_k v^t(\vq) A_k^a(\vp + \vq) \,. \label{Gl: Variationskern}
\end{align}
The Fourier transforms of the variational kernels $S$, $V$ and $W$ contained in our trial ansatz (\ref{Gl: VakuumfunktionalQ1}) are hereby given by
\begin{align}
S(p) &= \int \dd^3 x \, \exp(-\ii \vp \cdot \vx) S(x) \\
V(\vp, \vq) &= \int \dd^3 x \int \dd^3 y \, \exp(-\ii \vp \cdot \vx - \ii \vq \cdot \vy) V(\vx + \vz, \vy + \vz; \vz) \\
W(\vp, \vq) &= \int \dd^3 x \int \dd^3 y \, \exp(-\ii \vp \cdot \vx - \ii \vq \cdot \vy) W(\vx + \vz, \vy + \vz; \vz)
\end{align}
and are assumed to be scalar functions. In order to obey the correct behavior under spatial transformations, these kernels have to fulfill the relations $S(\vp) = S(p)$, $V(\vp, \vq) = V(-\vp, -\vq)$ and $W(\vp, \vq) = W(-\vp, -\vq)$, respectively. Note that contrary to the scalar kernel $S$, the vector kernels $V$ and $W$ have dimension of inverse momentum. In order to simplify our calculations, we will assume all kernels to be real valued functions and the vector kernels additionally to be symmetric, i.e. $V(\vp, \vq) = V(\vq, \vp)$ and $W(\vp, \vq) = W(\vq, \vp)$, respectively.

\section{Perturbative motivation of the ansatz for the fermionic vacuum state} \label{Anh: Stoerungstheorie}

In \cite{CR2015a}, QCD in Coulomb gauge was treated in Rayleigh--Schr\"odinger perturbation theory up to order $g^2$ in the coupling and the known results of covariant perturbation theory were reproduced. The perturbative results of ref.~\cite{CR2015a} yield a quark vacuum wave functional of the form of our ansatz (\ref{Gl: VakuumfunktionalQ}) and (\ref{Gl: VakuumfunktionalQ1}), however, with $S = 0$ and $W = 0$. The result obtained for the vector kernel $V$ agrees with the UV limit (\ref{Gl: VKernStoerungstheorie}) of our variational result (\ref{Gl: VKern}) as one expects due to asymptotic freedom. However, perturbation theory does not produce non-vanishing scalar and vector form factors, $S$ and $W$. As we have seen in the body of the paper both form factors are non-perturbative features. In particular, the variational result (\ref{Gl: WKern}) for $W$ disappears in the chirally symmetric  phase $S = 0$. In order to get information on the kinematic structure of the vector form factor $W$ we treat the quark-gluon coupling in Rayleigh--Schr\"odinger perturbation theory using, however, as unperturbed quark vacuum wave functional not the bare (perturbative) vacuum $\vert 0 \rangle$ but the BCS-state $\vert 0 \rangle_{\mathrm{BCS}}$ given by eq.~(\ref{Gl: VakuumfunktionalQ}) with $V = W = 0$ and $S \neq 0$ determined from the gap equation (\ref{Gl: Gapgleichung2}) with $\widetilde{g} = 0$. In this sense the BCS-wave functional represents an approximate solution of the functional Schr\"odinger equation for the Hamiltonian
\beq
\bar{H}'_{\mathrm{Q}} \equiv H_{\mathrm{Q}}^0 + H_{\mathrm{C}}^{\mathrm{Q}}
\eeq
in the mean-field approximation. What is left out in $\bar{H}'_{\mathrm{Q}}$ is the quark-gluon coupling $H_{\mathrm{Q}}^A$,
\beq
\bar{H}_{\mathrm{Q}} = \bar{H}'_{\mathrm{Q}}  + H_{\mathrm{Q}}^A \, ,
\eeq
whose effect we will now study in perturbation theory on top of the BCS-ground state $\vert \phi_{\mathrm{Q}}[A = 0] \rangle \equiv \vert 0 \rangle_{\mathrm{BCS}}$ which is in momentum space given by
\beq
\vert 0 \rangle_{\mathrm{BCS}} = \exp\left(-s \int \da^3 p \, S(p) {a^{s, m}}^{\dagger}(\vp) {b^{s, m}}^{\dagger}(-\vp)\right) \vert 0 \rangle \, , \label{Gl: BCSVakuum}
\eeq
see eqs.~(\ref{Gl: FermiAnsatz}) and (\ref{Gl: Variationskern}). Therefore, the ``bare'' QCD vacuum state which we consider now as starting point for perturbation theory reads $\vert 0 \rangle_{\mathrm{QCD}} = \vert 0 \rangle_{\mathrm{YM}} \otimes \vert 0 \rangle_{\mathrm{BCS}}$ where the bosonic part $\vert 0 \rangle_{\mathrm{YM}}$ is chosen similar to ref.~\cite{CR2015a}, i.e. by eq.~(\ref{Gl: BoseAnsatz1}) with $\omega(p)$ replaced by the photon energy $p$. The perturbative vacuum state is hence given by
\beq
\vert 0 \rangle_{\mathrm{pert}} \sim \vert 0 \rangle_{\mathrm{QCD}} + g \vert 0 \rangle_{\mathrm{QCD}}^1 + g^2 \vert 0 \rangle_{\mathrm{QCD}}^2 + \Oc(g^3) \label{Gl: VakuumStoerungstheorie}
\eeq
with the corrections $\vert 0 \rangle_{\mathrm{QCD}}^i$ being orthogonal to the unperturbed state $\vert 0 \rangle_{\mathrm{QCD}}$,
\beq
_{\mathrm{QCD}}\langle 0 \vert 0 \rangle_{\mathrm{QCD}}^i = 0 \quad \forall i \,.
\eeq
Rewriting the gauge fixed QCD Hamiltonian (\ref{Gl: QCDHamiltonian}) as a series in powers of the coupling $g$,
\beq
H_{\mathrm{QCD}} = H_0 + g H_1 + g^2 H_2 + \Oc(g^3) \,,
\eeq
the (first order) perturbative corrections to the vacuum $\vert 0 \rangle_{\mathrm{QCD}}$ is given by
\beq
\vert 0 \rangle_{\mathrm{QCD}}^1 = -\sum_N \frac{\langle N \vert H_1 \vert 0 \rangle_{\mathrm{QCD}}}{E_N - E_0} \vert N \rangle \label{Gl: Wellenfunktionalkorrektur}
\eeq
with $\vert N \rangle$ denoting a $N$-particle state with energy $E_N$ and $E_0$ being the (bare) vacuum energy. Note that the bare vacuum is not an eigenstate of the bare Hamiltonian $H_0$ (this would be given by $\vert 0 \rangle_{\mathrm{YM}} \otimes \vert 0 \rangle$) but of the mean field Hamiltonian given by the sum of $H_0$ and the (fermionic) effective single particle part of $H_2$ (the fermionic part of $g^2 H_2$ represents the Coulomb term $H_{\mathrm{C}}^{\mathrm{Q}}$ (\ref{Gl: CoulombtermQ})), as can be seen explicitly after performing a Bogoljubov transformation. In the quark sector, (\ref{Gl: Wellenfunktionalkorrektur}) gives the only relevant correction to the BCS-wave functional since the relevant contribution from $H_2$ is already included in the choice of the fermionic vacuum, eq.~(\ref{Gl: BCSVakuum}). The first order correction $H_1$ to the unperturbed Hamiltonian is given by the quark-gluon coupling
\beq
H_1 = \int \dd^3 x \, \psi^{\dagger}(\vx) \valpha \cdot \vA^a(\vx) t^a \psi(\vx) \label{Gl: Hamiltoniankorrektur} \,.
\eeq
Inserting the expansion (\ref{Gl: EntwicklungFeldop}) for the quark fields into (\ref{Gl: Hamiltoniankorrektur}), we find that the only non-vanishing contribution to the correction (\ref{Gl: Wellenfunktionalkorrektur}) comes from states containing one gluon, one quark and one anti-quark, i.e.
\beq
\vert N \rangle  \sim A_k^a(-\vp_1) {a^{s, m}}^{\dagger}(\vp_2) {b^{t, n}}^{\dagger}(\vp_3) \vert 0 \rangle_{\mathrm{QCD}}. \label{Gl: Korrekturansatz}
\eeq
With this wave function we can easily calculate the matrix element in the numerator of (\ref{Gl: Wellenfunktionalkorrektur}) yielding\footnote{The matrix element decomposes into one bosonic and one fermionic part. Both of them can be evaluated in the same way as the ones occuring in the body of this paper.}
\begin{align}
\langle N \vert H_1 \vert 0 \rangle_{\mathrm{QCD}} &\sim -t_{m n}^a t_{k l}(\vp_1) \frac{1}{2 p_1} \deltabar(\vp_1 + \vp_2 + \vp_3) \frac{1}{2 \sqrt{p_2 p_3}} P(p_2) P(p_3) \nonumber \\
&\phantom{\sim}\,\, \quad \times {u^{s}}^{\dagger}(\vp_2) \Bigl(\bigl[1 + S(p_2) S(p_3)\bigr] \alpha_l + \bigl[S(p_2) + S(p_3)\bigr] \beta \alpha_l\Bigr) v^{t}(\vp_3) \label{Gl: Korrektur}
\end{align}
where $t_{k l}(\vp) = \delta_{k l} - \hp_k \hp_l$ is the transversal projector in momentum space. From eq.~(\ref{Gl: Korrektur}) we can already read-off the (Dirac) structure of the corrections to the bare vacuum. Inserting (\ref{Gl: Korrektur}) together with (\ref{Gl: Korrekturansatz}) into (\ref{Gl: Wellenfunktionalkorrektur}) we find for the fermionic part of the (perturbative) wave functional
\beq
\vert 0 \rangle_{\mathrm{pert}}^{\mathrm{Q}} \sim \left(1 + g \int \dd^3 x \int \dd^3 y \int \dd^3 z \, \psi_+^{\dagger}(\vx) \Bigl[f(\vx, \vy; \vz) + g(\vx, \vy; \vz) \beta\Bigr] \valpha \cdot \vA^a(\vz) t^a \psi_-(\vy)\right) \vert 0 \rangle_{\mathrm{BCS}} \label{Gl: Wellenfunktionalkorrektur1}
\eeq
where we have switched to the coordinate space representation for sake of comparison with our ansatz (\ref{Gl: VakuumfunktionalQ}). Equation (\ref{Gl: Wellenfunktionalkorrektur1}) exhibits precisely the (Dirac) structure of the quark-gluon coupling assumed in our ansatz (\ref{Gl: VakuumfunktionalQ}) for the vacuum wave functional. Let us also mention that the $S$-dependence of the numerator of (\ref{Gl: Korrektur}) perfectly agrees with the one obtained for the vector kernels from the variational principle, see eqs.~(\ref{Gl: VKern}) and (\ref{Gl: WKern}). Note, furthermore, that the results obtained here reduce to the ones of ref.~\cite{CR2015a} when the limit $S \to 0$ is considered. Particularly, the additional term with the Dirac structure $\beta \alpha_i$ vanishes in this limit and can therefore not be recovered in perturbation theory starting from the bare fermionic vacuum $\vert 0 \rangle$.

\section{Explicit calculation of the static quark propagator} \label{Anh: Propagator}

Using the fermionic anticommutator (\ref{Gl: Antikommutator}) and the expansion (\ref{Gl: EntwicklungFeldop}) of the quark field, the static quark propagator (\ref{Gl: StatProp}) can be expressed as
\begin{align}
&G_{i j}^{m_1 m_2}(\vx, \vy) + \frac{1}{2} \delta^{m_1 m_2} \delta_{i j} \delta(\vx - \vy) = \langle \psi_i^{m_1}(\vx) {\psi_j^{m_2}}^{\dagger}(\vy) \rangle = \nonumber \\
&\quad= \int \da^3 p_1 \int \da^3 p_2 \, \frac{1}{2 \sqrt{p_1 p_2}} \Bigl(\langle a^{s_1, m_1}(\vp_1) {a^{s_2, m_2}}^{\dagger}(\vp_2) \rangle u_i^{s_1}(\vp_1) {u_j^{s_2}}^{\dagger}(\vp_2) \exp\bigl(\ii \vp_1 \cdot \vx - \ii \vp_2 \cdot \vy\bigr) \nonumber \\
&\quad\phantom{=}\,\, \phantom{\int \da^3 p_1 \int \da^3 p_2 \, \frac{1}{2 \sqrt{p_1 p_2}} \Bigl(} + \langle a^{s_1, m_1}(\vp_1) b^{s_2, m_2}(\vp_2) \rangle u_i^{s_1}(\vp_1) {v_j^{s_2}}^{\dagger}(\vp_2) \exp\bigl(\ii \vp_1 \cdot \vx + \ii \vp_2 \cdot \vy\bigr) \nonumber \\
&\quad\phantom{=}\,\, \phantom{\int \da^3 p_1 \int \da^3 p_2 \, \frac{1}{2 \sqrt{p_1 p_2}} \Bigl(} + \langle {b^{s_1, m_1}}^{\dagger}(\vp_1) {a^{s_2, m_2}}^{\dagger}(\vp_2) \rangle v_i^{s_1}(\vp_1) {u_j^{s_2}}^{\dagger}(\vp_2) \exp\bigl(-\ii \vp_1 \cdot \vx - \ii \vp_2 \cdot \vy\bigr) \nonumber \\
&\quad\phantom{=}\,\, \phantom{\int \da^3 p_1 \int \da^3 p_2 \, \frac{1}{2 \sqrt{p_1 p_2}} \Bigl(} + \langle {b^{s_1, m_1}}^{\dagger}(\vp_1) b^{s_2, m_2}(\vp_2) \rangle v_i^{s_1}(\vp_1) {v_j^{s_2}}^{\dagger}(\vp_2) \exp\bigl(-\ii \vp_1 \cdot \vx + \ii \vp_2 \cdot \vy\bigr)\Bigr) \label{Gl: StatProp2} \,.
\end{align}
The fermionic expectation values of the quark two-point functions were given in eq.~(\ref{Gl: Operatorpaerchen}) in coordinate space. After Fourier transformation one finds e.g.
\beq
\langle a^{s_1, m_1}(\vp_1) {a^{s_2, m_2}}^{\dagger}(\vp_2) \rangle_{\mathrm{Q}} = {\left[\Id + K K^{\dagger}\right]^{-1}}^{s_1 s_2, m_1 m_2}(\vp_1, \vp_2) \,.
\eeq
The bosonic expectation value of this quantity is taken by using eq.~(\ref{Gl: Wick2}),
\begin{align}
\langle a^{s_1, m_1}(\vp_1) {a^{s_2, m_2}}^{\dagger}(\vp_2) \rangle &= {\left[\Id + K K^{\dagger}\right]^{-1}}^{s_1 s_2, m_1 m_2}(\vp_1, \vp_2) \Bigr\vert_{A = 0} \nonumber \\
&\phantom{=}\,\, + \frac{1}{4} \int \da^3 q \, \frac{\delta}{\delta A_k^a(\vq)} t_{k l}(\vq) \frac{1}{\omega(q)} \frac{\delta}{\delta A_l^a(-\vq)} {\left[\Id + K K^{\dagger}\right]^{-1}}^{s_1 s_2, m_1 m_2}(\vp_1, \vp_2)\Bigr\vert_{A = 0} \,.
\end{align}
With
\beq
(K K^{\dagger})^{s_1 s_2, m_1 m_2}(\vp_1, \vp_2) \Bigr\vert_{A = 0} = \delta^{s_1 s_2} \delta^{m_1 m_2} \deltabar^3(\vp_1 - \vp_2) S^2(p_1) \,,
\eeq
we find for the leading order contribution
\beq
{\left[\Id + K K^{\dagger}\right]^{-1}}^{s_1 s_2, m_1 m_2}(\vp_1, \vp_2) \Bigr\vert_{A = 0} = \delta^{s_1 s_2} \delta^{m_1 m_2} \deltabar^3(\vp_1 - \vp_2) P(p_1) \,.
\eeq
For the next to leading order contribution, we obtain after a short calculation
\begin{align}
&\frac{\delta}{\delta A_k^a(\vq)} \frac{\delta}{\delta A_l^a(-\vq)} {\left[\Id + K K^{\dagger}\right]^{-1}}^{s_1 s_2, m_1 m_2}(\vp_1, \vp_2)\Bigr\vert_{A = 0} = \nonumber \\
&\quad = P(p_1) P(p_2) \int \da^3 p_3 \, P(p_3) \left.\frac{\delta (K K^{\dagger})^{s_1 s_3, m_1 m_3}(\vp_1, \vp_3)}{\delta A_k^a(\vq)} \frac{\delta (K K^{\dagger})^{s_3 s_2, m_3 m_2}(\vp_3, \vp_2)}{\delta A_l^a(-\vq)}\right\vert_{A = 0} \nonumber \\
&\quad \phantom{=}\,\, + P(p_1) P(p_2) \int \da^3 p_3 \, P(p_3) \left.\frac{\delta (K K^{\dagger})^{s_1 s_3, m_1 m_3}(\vp_1, \vp_3)}{\delta A_l^a(-\vq)} \frac{\delta (K K^{\dagger})^{s_3 s_2, m_3 m_2}(\vp_3, \vp_2)}{\delta A_k^a(\vq)}\right\vert_{A = 0} \nonumber \\
&\quad \phantom{=}\,\, - P(p_1) P(p_2) \frac{\delta^2 (K K^{\dagger})^{s_1 s_2, m_1 m_2}(\vp_1, \vp_2)}{\delta A_k^a(\vq) \delta A_l^a(-\vq)} \,.
\end{align}
Inserting here for the variational kernel the explicit form (\ref{Gl: Variationskern}) and carrying out the functional derivatives $\delta / \delta A$ the Dirac and color structure of these expressions can be worked out using the relations of appendix \ref{Anh: Impulsdarstellung}. One finds eventually
\beq
\langle a^{s_1, m_1}(\vp_1) {a^{s_2, m_2}}^{\dagger}(\vp_2) \rangle = \delta^{s_1 s_2} \delta^{m_1 m_2} \deltabar^3(\vp_1 - \vp_2) P(p_1) \left[1 - \frac{1}{2} I_{\alpha}(\vp_1)\right] \label{Gl: Operatorpaerchen1}
\eeq
where $I_{\alpha}$ is defined in eq.~(\ref{Gl: Schleifenintegralalpha}). Inserting now eq.~(\ref{Gl: Operatorpaerchen1}) back into eq.~(\ref{Gl: StatProp2}) eventually enables us to calculate the outer product $u u^{\dagger}$ of the eigenspinors yielding an expression given in terms of Dirac matrices, see appendix \ref{Anh: Impulsdarstellung}. Performing a similar calculation for the remaining two-point functions in (\ref{Gl: StatProp2}) yields finally the expression given in eq.~(\ref{Gl: StatProp1}).

\section{Ground state energy revisited} \label{Anh: EW}

\subsection{Expectation value of the free Dirac Hamiltonian}

Inserting the explicit expression for the quark propagator (\ref{Gl: StatProp1}) into eq.~(\ref{Gl: FrDiracHamEW2}) and taking the trace yields the following expression for the expectation value of the free Dirac Hamiltonian:
\begin{align}
\langle H_{\mathrm{Q}}^0 \rangle &= 2 \Ncc \deltabar^3(0) \int \da^3 p \, p \bigl(1 - 2 P(p)\bigr) \nonumber \\
&\phantom{=}\,\, + \frac{\Ncc^2 - 1}{2} \deltabar^3(0) g^2 \int \da^3 p \int \da^3 q \, \frac{V^2(\vp, \vq)}{\omega(|\vp + \vq|)} P(p) P(q) X(\vp, \vq) \nonumber \\
&\phantom{=}\,\, \quad \quad \times \Bigl[p P(p) \bigl(1 - S^2(p) + 2 S(p) S(q)\bigr) + q P(q) \bigl(1 - S^2(q) + 2 S(p) S(q)\bigr)\Bigr] \nonumber \\
&\phantom{=}\,\, + \frac{\Ncc^2 - 1}{2} \deltabar^3(0) g^2 \int \da^3 p \int \da^3 q \, \frac{W^2(\vp, \vq)}{\omega(|\vp + \vq|)} P(p) P(q) Y(\vp, \vq) \nonumber \\
&\phantom{=}\,\, \quad \quad \times \Bigl[p P(p) \bigl(1 - S^2(p) - 2 S(p) S(q)\bigr) + q P(q) \bigl(1 - S^2(q) - 2 S(p) S(q)\bigr)\Bigr] \label{Gl: FrDiracHamEW1}
\end{align}

\subsection{Expectation value of the kinetic energy}

In order to calculate the expectation value of $\widetilde{H}_{\mathrm{YM}}^E$ (\ref{Gl: TransfKinEn}), we consider first those terms of $\widetilde{H}_{\mathrm{YM}}^E$ which explicitly contain the Faddeev--Popov determinant $J$. Using the approximation (\ref{Gl: FaddeevPopov1}), the functional derivatives induced by the canonical momentum operator $\Pi$ can be easily carried out yielding expressions whose fermionic expectation value is trivial while the bosonic one can be calculated in terms of Wick's theorem (\ref{Gl: Wick1}) leading to the result
\begin{subequations}
\begin{align}
-\frac{1}{8} \Bigl\langle \int \da^3 p \, \bigl[\Pi_k^a(\vp) \ln J \bigr] \bigl[\Pi_k^a(-\vp) \ln J \bigr] \Bigr\rangle &= \frac{\Ncc^2 - 1}{2} \deltabar^3(0) \int \da^3 p \, \frac{\chi^2(p)}{\omega(p)} \\
-\frac{1}{4} \Bigl\langle \int \da^3 p \, \bigl[\Pi_k^a(\vp) \Pi_k^a(-\vp) \ln J \bigr] \Bigr\rangle &= -(\Ncc^2 - 1) \delta^3(0) \int \da^3 p \, \chi(p) \,.
\end{align}
\label{Gl: EWKinEn1}%
\end{subequations}
For the evaluation of the expectation value of those parts of $\widetilde{H}_{\mathrm{YM}}^E$ (\ref{Gl: TransfKinEn}) which do not contain the Faddeev--Popov determinant $J$ it is convenient to carry out an integration by parts,\footnote{Notice that this is only possible in the first two terms on the r.h.s.~of eq.~(\ref{Gl: TransfKinEn}).}
\beq
\int \Dc A \, \widetilde{\phi}_{\mathrm{YM}}^*[A] f[A] \Pi \, g[A] \widetilde{\phi}_{\mathrm{YM}}[A] = -\int \Dc A \, \left[\Pi \, \widetilde{\phi}_{\mathrm{YM}}^*[A] f[A]\right] g[A] \widetilde{\phi}_{\mathrm{YM}}[A] \,.
\eeq
From the expectation value of the second term on the r.h.s.~of eq.~(\ref{Gl: TransfKinEn}) one obtains then the relation
\beq
2 \int \da^3 p \int \Dc A \, \widetilde{\phi}_{\mathrm{YM}}[A] \Bigl[\Pi_i^a(\vp) \ln I[A] \Bigr] \left[\Pi_i^a(-\vp) \widetilde{\phi}_{\mathrm{YM}}[A]\right] = -\int \da^3 p \int \Dc A \, \widetilde{\phi}_{\mathrm{YM}}^2[A] \Bigl[\Pi_i^a(-\vec{p}) \Pi_i^a(\vec{p}) \ln I[A] \Bigr]
\eeq
where $I[A] = \langle \phi_{\mathrm{Q}}[A] \vert \phi_{\mathrm{Q}}[A] \rangle$ and $\widetilde{\phi}_{\mathrm{YM}} = \widetilde{\phi}_{\mathrm{YM}}^*$ (which follows directly from our ansatz (\ref{Gl: BoseAnsatz1}) for a real valued kernel $\omega$) have been used. With the help of this relation it is possible to write the expectation value of those parts of $\widetilde{H}_{\mathrm{YM}}^E$ (\ref{Gl: TransfKinEn}) which do not contain $J$ in the form
\begin{align}
\langle H_{\mathrm{YM}}^E \rangle\bigr\vert_{J = 0} &= -\frac{1}{2} |\Nc|^2 \int \da^3 p \int \Dc A \, \left[\Pi_k^a(\vp) \widetilde{\phi}_{\mathrm{YM}}[A]\right] \left[\Pi_k^a(-\vp) \widetilde{\phi}_{\mathrm{YM}}[A]\right] \nonumber \\
&\phantom{=}\,\, - \frac{1}{2} |\Nc|^2 \int \da^3 p \int \Dc A \, \widetilde{\phi}_{\mathrm{YM}}^2[A] I^{-1}[A] \Bigl[\Pi_k^a(\vp) \langle \phi_{\mathrm{Q}}[A] \vert\Bigr] \Bigl[\Pi_k^a(-\vp) \vert \phi_{\mathrm{Q}}[A] \rangle\Bigr] \nonumber \\
&\phantom{=}\,\, + \frac{1}{8} |\Nc|^2 \int \da^3 p \int \Dc A \, \widetilde{\phi}_{\mathrm{YM}}^2[A] \Bigl[\Pi_k^a(\vp) \ln I[A] \Bigr] \Bigl[\Pi_k^a(-\vp) \ln I[A] \Bigr] \,. \label{Gl: EWKinEn}
\end{align}
The further evaluation of this is straightforward: After carrying out the functional derivatives induced by $\Pi$, the (purely bosonic) first term on the r.h.s. of eq.~(\ref{Gl: EWKinEn}) can be expressed in terms of the gluon propagator (\ref{Gl: Gluonpropagator}) while the (fermionic) second and third terms yield the product of four quark fields which can be treated similar to the expectation value of the fermionic part $H_{\mathrm{C}}^{\mathrm{Q}}$ of the Coulomb term, see section \ref{Abschn: EW} in the body of the paper. Finally, we obtain for the expectation value of those terms \textit{not} containing the Faddeev--Popov determinant
\begin{align}
\langle H_{\mathrm{YM}}^E \rangle\bigr\vert_{J = 0} &= \frac{\Ncc^2 - 1}{2} \deltabar^3(0) \int \da^3 p \, \omega(p) \nonumber \\
&\phantom{=}\,\, + \frac{\Ncc^2 - 1}{2} \deltabar^3(0) g^2 \int \da^3 p \int \da^3 q \, P(p) P(q) V^2(\vp, \vq) X(\vp, \vq) \nonumber \\
&\phantom{=}\,\, + \frac{\Ncc^2 - 1}{2} \deltabar^3(0) g^2 \int \da^3 p \int \da^3 q \, P(p) P(q) W^2(\vp, \vq) Y(\vp, \vq) \,.
\end{align}
If we add now (\ref{Gl: EWKinEn1}) to this expression, we end up with the result given in section \ref{Abschn: EW}, eq.~(\ref{Gl: Kinen1}).

\section{UV expansion} \label{Anh: UV}

\subsection{Quark gap equation}

For expanding the integrands on the r.h.s. of eq.~(\ref{Gl: Gapgleichung}) in a Taylor series we need an assumption for the UV behavior of the gluon energy $\omega$ and the scalar kernel $S$. For the first we can conclude immediately from Gribov's formula (\ref{Gl: Gribov}) $\omega(p \to \infty) = p$ while for the latter $S(p\to \infty) \to 0$ should hold because of asymptotic freedom (in previous work a behavior of $S \sim p^{-5}$ was obtained in the limit of a vanishing quark-gluon coupling \cite{Adler1984, Pak2013, Watson2012}). Therefore, we will assume the scalar kernel to vanish sufficiently fast for not being involved in any UV divergence, i.e. $S(p) \approx 0$. Note that we will use the UV form of $S$ and $\omega$ only for functions depending on the loop momentum $p$.

As an illustration, in the following we demonstrate the basic steps for the UV analysis of the fourth integral on the r.h.s. of eq.~(\ref{Gl: Gapgleichung}), $I_{V \mathrm{Q}}^{\mathrm{Q}}$ [eq.~(\ref{Gl: KopplungstermGapglV})]. Introducing $\xi = k / p \ll 1$ and $z = \hp \cdot \hk$, we find for the vector kernel $V$ [eq.~(\ref{Gl: VKern})] the expansion
\begin{align}
V(\vp, \vk) &= \frac{1 + S(p) S(k)}{p P(p) \Bigl(1 - S^2(p) + 2 S(p) S(k)\Bigr) + k P(k) \Bigl(1 - S^2(k) + 2 S(p) S(k)\Bigr) + \omega(|\vp + \vk|)} \nonumber \\
&\approx \Bigl[p + k P(k) \Bigl(1 - S^2(k)\Bigr) + \omega(|\vp + \vk|)\Bigr]^{-1} \nonumber \\
&\approx \left[p + k P(k) \Bigl(1 - S^2(k)\Bigr) + \Bigl(p^2 + k^2 + 2 p k z\Bigr)^{1/2}\right]^{-1} \nonumber \\
&\approx \frac{1}{2 p} \left[1 - \frac{1}{2} \xi \Bigl(P(k) \Bigl(1 - S^2(k)\Bigr) + z\Bigr) + \Oc(\xi^2)\right]
\end{align}
and in a similar fashion for the remaining contributions
\begin{align}
\frac{1}{\omega(|\vp + \vk|)} &= \frac{1}{p} \Bigl[1 - \xi z + \Oc(\xi^2)\Bigr] \\
X(\vp, \vk) &= 1 - z - \xi \bigl(1 - z^2\bigr) + \Oc(\xi^2) \,.
\end{align}
After inserting this into the integral (\ref{Gl: KopplungstermGapglV}) we end up with\footnote{We consider the integral only up to the cutoff $\Lambda/\mu$.}
\begin{align}
I_{V \mathrm{Q}}^{\mathrm{Q}}(k) &\approx -\frac{C_{\mathrm{F}} \mu}{2 (2 \pi)^3} g^2 S(k) \int\limits^{\Lambda/\mu} \dd p \int\limits_{-1}^1 \dd z \int\limits_0^{2 \pi} \dd \varphi_{p} \left\{1 - z - \xi \left[1 + \frac{3}{2} z - \frac{5}{2} z^2 + \frac{1}{2} P(k) \Bigl(1 - S^2(k)\Bigr) (1 - z)\right]\right\} \nonumber \\
&= -\frac{C_{\mathrm{F}}}{8 \pi^2} g^2 S(k) \left\{2 \Lambda - k \left[\frac{1}{3} + P(k) \Bigl(1 - S^2(k)\Bigr)\right] \ln \frac{\Lambda}{\mu} + \Oc(1/\Lambda)\right\}.
\end{align}

Performing the same analysis for the other integrals on the r.h.s. of the gap equation (\ref{Gl: Gapgleichung}) we find for the contributions containing the vector kernel $V$ the (divergent) UV behavior
\beq
\frac{C_{\mathrm{F}}}{16 \pi^2} g^2 S(k) \left[-2 \Lambda + k \ln \frac{\Lambda}{\mu} \left(-\frac{2}{3} + 4 P(k)\right)\right]
\eeq
and for those containing the vector kernel $W$
\beq
\frac{C_{\mathrm{F}}}{16 \pi^2} g^2 S(k) \left[2 \Lambda + k \ln \frac{\Lambda}{\mu} \left(\frac{10}{3} - 4 P(k)\right)\right].
\eeq
Note that in the sum of these two terms the linear divergence exactly cancels. Adding finally the UV divergence stemming from the UV part of the Coulomb kernel (\ref{Gl: Coulombkern2}),
\beq
\frac{C_{\mathrm{F}}}{6 \pi^2} g^2 k S(k) \ln \frac{\Lambda}{\mu} \,,
\eeq
we end up with the divergent factor given in eq.~(\ref{Gl: UVDivGapgl}).

\subsection{Quark propagator}

Analogously to the UV analysis of the gap equation one can also consider the UV limit of the loop integrals $I_{\alpha}$ [eq.~(\ref{Gl: Schleifenintegralalpha})] and $I_{\beta}$ [eq.~(\ref{Gl: Schleifenintegralbeta})], contained in the static quark propagator (\ref{Gl: StatProp1}). One finds the (logarithmically divergent) UV behavior
\begin{align}
I_{\alpha}(p) &= \frac{C_{\mathrm{F}} g^2}{8 \pi^2} \Bigl(1 - S^2(p)\Bigr) \ln \frac{\Lambda}{\mu} + \text{finite terms} \label{Gl: StatPropDiv1} \\
I_{\beta}(p) &= \frac{C_{\mathrm{F}} g^2}{8 \pi^2} S(p) \ln \frac{\Lambda}{\mu} + \text{finite terms} \,. \label{Gl: StatPropDiv2}
\end{align}

\bibliography{QCDT0}

\end{document}